%% file: article.tex
\pdfoutput=1
\documentclass[a4paper,DIV15,10pt,ngerman,twocolumn]{scrartcl}
\usepackage[utf8]{inputenc}
\usepackage[T1]{fontenc} 
\usepackage{graphicx}
\usepackage{authblk}

\usepackage{amsmath}
\usepackage{gensymb}

\usepackage{epstopdf}
\usepackage[toc,page]{appendix}

\usepackage[pdftex,
            pdfauthor={Anton Gladky, Rüdiger Schwarze},
            pdftitle={Comparison of different capillary bridge models for application in the discrete element method},
            pdfsubject={Granular material, DEM},
            pdfkeywords={Granular material,  DEM, Capillary bridge models, Liquid bridges},
            pdfcreator={pdflatex}]{hyperref}
            
\begin{document}

\title{Comparison of different capillary bridge models for application in the discrete element method}

\makeatletter
\author[*]{Anton Gladkyy}
\author[*]{Rüdiger Schwarze}
\affil[*] {
  Institute of Mechanics and Fluid Dynamics, TU Bergakademie Freiberg,
  Lampadiusstr. 4, 09596 Freiberg, Germany.
}

\maketitle
\input{main.tex}

\bibliographystyle{acm}      
\bibliography{capillar}   

\input{appendix.tex}

\end{document}

%% file: main.tex
\begin{abstract}
Weakly wetted granular material is the subject of many studies. Several formulations
were proposed  to calculate the capillary forces between wet particles. 
In this paper some of such models have been implemented in a DEM-framework, and simulation
results were compared to experimental measurements. Also, the influence
of capillary model type on macro parameters like local shear viscosity
and cohesive parameters of sheared material have been investigated
through the simulation of spherical beads using a  DEM-model of a split-bottom
shear-cell.

It was concluded that the water content, simulated with the help of
capillary bridge models, changes the macro-properties of the simulated granular
material. Different capillary bridge models do not influence the
macroscopic results visibly.

\end{abstract}

\section{Introduction}
  \label{intro}
Wet granular materials play an important role in geology and many technical 
applications, e.g. construction, pharmaceuticals, civil engineering, etc. 
Here, liquid or capillary bridges are present between the grains, 
which produce inter-grain forces on the micro-scale level and drastically modify the
mechanical properties of the granular media on the macro-scale levels (e.g. slopes
can be much larger than 45 degrees for banked up wet bulk material)
\cite{Iveson2001Nucleation,Herminghaus2005Dynamics,Mitarai2006Wet}.

Such inter-grain capillary forces have been the subject of many investigations 
within the last decades, see e.g. 
\cite{Lambert2004CapForce,Herminghaus2005Dynamics,lambert2008comparison} for an 
overview. Modern instruments allow one to measure micro- and even nano-scale force values 
between individual particles very precisely. Based on this experimental information, 
capillary bridge models (CBMs) for capillary force calculations can be deduced, as 
described e.g. in \cite{Lambert2007CapBook}. Some recent studies in this direction 
have been done for instance by Willett et al. \cite{Willett2000} and Rabinovich et al.
\cite{Rabinov2005}.
  
These CBMs can be implemented in particle-based simulation methods, for example on the base of the
Discrete Element Method (DEM), in order to model the effects of individual capillary 
bridges between particles. Using a CBM it is possible to simulate the 
behavior of wetted bulk material and to predict its macro-parameters. This is 
important for simulations of processes like agglomeration, adhesion, crystallization 
and others \cite{Scholtes2009SoilDEM,Radl2010MixingDEM,Liu2011DrumDEM}. 
  
In this paper, we investigate four different CBMs, which have been 
frequently used to model capillary bridge forces in DEM simulations. Firstly, 
the CBM formulations and their implementations in an open-source DEM 
software package are summarized. Then, the quality of the CBM for the 
description of individual capillary bridges in the pendular regime (where the 
capillary bridges exist individually) is tested by comparison of experimental 
and numerical data. Finally, every CBM is analyzed in simulations of 
weakly wetted bulk material shear flows by cross-comparison of 
results from the different CBMs. Here, the bridges are again in pendular state.

\section{Employed capillary bridge models}
\label{implement}

Zhu et al. \cite{zhu07} discuss in their review the main features of CBMs for the
calculation of capillary forces. CBMs for application in DEM simulations should 
consist of easy to implement explicit functions of liquid bridge volume and the 
particle separation distance. Additionally, the shape of the liquid bridge must 
be approximated in order to calculate the capillary force. 

The total capillary attractive force between two particles is caused by a surface
tension component and hydrostatic pressure in the bulk \cite{lian93}.

Two different methods can be adopted: in the neck (or ``gorge'') method the force is
estimated at the neck of the bridge. In contrast, in the contact (or boundary) method,
the force is evaluated at the liquid bridge solid contact region. It has been
demonstrated that both methods provide relatively precise predictions of capillary
forces \cite{zhu07}.

Some recent approaches with explicit CBM functions have been proposed by Weigert 
and Ripperger \cite{Weigert1999} (contact method), Willett et al. \cite{Willett2000} 
(neck method), Rabinovich et al. \cite{Rabinov2005} (neck method) and Soulie et al.
\cite{soul06} (neck method). Lambert et al. \cite{lambert2008comparison} give a 
corrected version of Rabinovich's CBM. Several of these CBMs have already been
successfully used in DEM simulations of weakly wetted granular material 
\cite{Radl2010MixingDEM,gabr12,gan13,mani13,Schwarze2013a}. A comparison of 
CBMs from Soulie and Rabinovich (with Lambert's correction) has already been given in 
\cite{gabr12}. Therefore, we will compare all stated CBMs except Soulie's in order to 
identify (i) differences between CBM based on the contact and the neck method and (ii) 
differences between CBMs based on the same method (neck) but with different formulations.

\begin{figure}[h!]
  \begin{center}
     \includegraphics[width=1.00\linewidth]{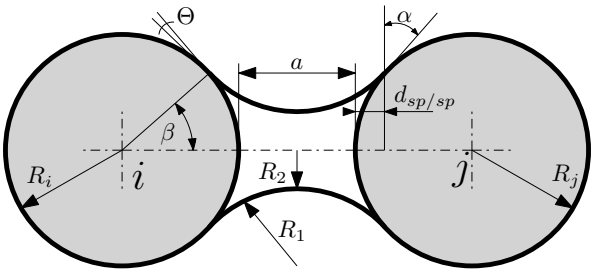}
  \end{center}
  \caption{Pendular liquid bridge between two spherical particles. See the text for
  the description of the indicated parameters.}
  \label{fig:Schemacapillar}
\end{figure}

In our study, we apply four different CBMs, which have been proposed in chronological 
order by Weigert and Ripperger \cite{Weigert1999}, Willett et al. \cite{Willett2000} 
and Rabinovich et al. \cite{Rabinov2005}) for the capillary force $F_{cap}$ between 
particles $i$  and $j$ with radii $R_i$ and $R_j$. In these models, an effective radius 
$R$ of the particle pair (which is based on the Derjaguin approximation) is calculated
as a harmonic mean of 2 different radii.
      \begin{equation}
        \label{eq:Rreduced}
        R = \cfrac{2 R_i R_j}{R_i + R_j}
      \end{equation}

Willett et al. \cite{Willett2000} have demonstrated that the Derjaguin approximation works in a reasonable way for separation distances
$a$ excluding close-contact ($a \approx 0$) and near rupture distance
($a \approx a_{crit}$). Some limitations of this approach have been 
discussed  by Rabinovich et al. \cite{Rabinov2005}. 

As constant input parameters, we prescribe the liquid bridge volume $V=V_{bridge}$, 
the contact angle $\theta$ of the liquid-gas interface at the particle surface, and 
the liquid surface tension $\gamma$. In all four CBMs, the distance $a$ between the 
surfaces of the two particles $i$, $j$ is the main model variable
(Fig. \ref{fig:Schemacapillar}). We assume that $F_{cap}$
(capillary force) is acting only after a mechanical contact between $i$ and $j$.
$F_{cap}$  is acting as long as $a < a_{crit}$.
The critical distance (rupture distance of the liquid bridge) $a_{crit}$
between two particles is calculated according to 
Willett et al. \cite{Willett2000}:
  \begin{equation}
    \label{eq:Scrit}
    a_{crit}^{*} = \left( 1 + \cfrac{\theta}{2}\right) \cdot \left( {V^{*}}^{1/3} + \cfrac{{V^{*}}^{2/3}}{10}\right)
  \end{equation}
Where $a_{crit}^*=a_{crit}/R$ and $V^{*}=V/R^{3}$ are the dimensionless
critical distance and bridge volume accordingly.

All formulations, which are given below, have been implemented in the open-source 
DEM software Yade \cite{yade:doc} in combination with a linear viscoelastic model for 
normal contacts \cite{radjai2011discrete}:
  \begin{equation}
    \label{eq:Fviscoel}
    \left.
      \begin{array}{l l}
        F_{norm} &= -k_{n} \delta_{n} - c_{n}\cfrac{\textrm{d}\delta_{n}}{\textrm{dt}} \\
        F_{tan}  &= -k_{t} \delta_{t} - c_{t}\cfrac{\textrm{d}\delta_{t}}{\textrm{dt}}
      \end{array}
    \right.
  \end{equation}
Here, spring and damping parameters $k_{n}$, $k_{t}$, $c_{n}$ and $c_{t}$
are calculated as a function of restitution coefficients in normal and 
tangential directions $e_{n}$ and $e_{t}$, contact duration time $t_c$ and mass of particles,
see \cite{Pournin2001} for details. The leapfrog algorithm is used for the integration of motion.
Our model implementation is in line with the description given in 
\cite{radjai2011discrete}.

The total interaction force acting on particle during the normal contact
is the sum of normal and tangential forces  Eq. (\ref{eq:Ftotal}). Coulomb's
friction law  $|F_{tan}| \le \mu_{C} F_{norm}$ as well as preventing undesirable
attractive forces when $\textrm{d}\delta_{n}/\textrm{dt} > 0$
were considered in implementation  \cite{Schwager2007}. Rolling resistance is neglected.
The capillary force $\vec{F}_{cap}$ is acting along the line connecting the
centers of particles $j$ and $i$.

  \begin{equation}
    \label{eq:Ftotal}
    \vec{F} =
    \left\{
      \begin{array}{l l}
         \vec{F}_{norm} + \vec{F}_{tan}, & \delta_{n}<0 \\
         \vec{F}_{cap},                  & \delta_{n}>0 
      \end{array}
    \right.
  \end{equation}

We are neglecting effects of capillary forces during mechanical
contact because we are not aware of a proved model for that case.
Because the capillary forces are in order of magnitude much smaller than
the mechanical ones, we do not expect a large influence from this
simplification.

\subsection{Weigert's model}
\label{weigMod}
In the CBM of Weigert and Ripperger \cite{Weigert1999}, empirical equations are 
employed in order to calculate $V$ and the half-filling angle $\beta$ from the bulk 
liquid saturation. These parameters are needed to calculate $F_{cap}$:
      \begin{equation}
        \label{eq:WEI:21}
        F_{cap} = \underbrace{\cfrac{\pi}{4} ({2R})^2 \sin^2{\beta} \cdot  p_k}_{F_p} + \underbrace{\gamma \pi 2R \sin{\beta} \sin{(\beta + \theta)}}_{F_{\gamma}}
      \end{equation}
where the first term $F_{p}$ is the hydrostatic pressure component and the second one $F_{\gamma}$
is the liquid surface tension contribution.
Both force components are evaluated at the particle surface.
Main geometrical parameters of the liquid bridge are
showed in figure \ref{fig:Schemacapillar}.

Equations for all other parameters are presented in the appendix.
The sequence of equations in implementation is the following: Eq. (\ref{eq:WEI:16}),
Eq. (\ref{eq:WEI:17}), Eq. (\ref{eq:WEI:15Mod}), Eq. (\ref{eq:Piet:2}), Eq. (\ref{eq:Piet:3}), 
Eq. (\ref{eq:WEI:22}), Eq. (\ref{eq:WEI:21}).

\subsection{Willett's full model}
\label{wilNMod}
Willett et al. \cite{Willett2000} proposed the following CBM, which is based on 
combined experimental and numerical results with numerical data from integration of 
the Laplace-Young equation. The main model variable is the scaled, dimensionless half-
separation distance $S^{+}$:
      \begin{equation}
        \label{eq:WIL:11}
         S^{+} =  \cfrac{a}{2 \sqrt{V / R}}
      \end{equation}
$F_{cap}$ is calculated from:
  \begin{equation}
  \label{eq:WIL:A1-lnFn-Mod1}
   F_{cap} = 2\,\pi R \gamma\,\exp({f_1 - f_2 \exp(f_3 \ln{S^{+}} + f_4 {\ln^2{S^{+}}})})
  \end{equation}

The definitions of the coefficients $f_{1\dots 4}$ are presented in the appendix
and were derived by curve-fitting to the numerical solution.
The sequence of equations in implementation is the following:
    Eq. (\ref{eq:WIL:A1-f1}), Eq. (\ref{eq:WIL:A1-f2}), Eq. (\ref{eq:WIL:A1-f3}), 
    Eq. (\ref{eq:WIL:A1-f4}), Eq. (\ref{eq:WIL:11}), Eq. (\ref{eq:WIL:A1-lnFn-Mod1}).

\subsection{Willett's reduced model}
\label{wilAMod}    
Willett et al. \cite{Willett2000} also give a less complex CBM for equal-sized 
particles $R_i = R_j = R$. The following equation provides a closed approximation of 
$F_{cap}$ between equal-sized spheres:
      \begin{equation}
        \label{eq:WIL:12}
        \begin{split}
          F_{cap}= \cfrac{2\,\pi R \gamma\,\cos{\theta}}{1 + 2.1({S^{+}}) + 10 ({S^{+}})^2}
        \end{split}
      \end{equation}
The sequence of equations in implementation is the following: Eq. (\ref{eq:WIL:11}), Eq. (\ref{eq:WIL:12}).

Willett  et al. \cite{Willett2000} noticed that both proposed formulations are valid for 
$\theta < 50 \degree$ and $V^{*} < 0.1$. 

\subsection{Rabinovich's model}
\label{rabiMod}
Rabinovich et al. \cite{Rabinov2005} give the following CBM, which is based on 
combined experimental and numerical analysis as well. First of all, the ``embracing angle'' 
$\alpha$ for the case sphere-sphere is evaluated (see Fig. \ref{fig:Schemacapillar}):
      \begin{equation}
        \label{eq:RAB:A3}
        \alpha = \sqrt{\cfrac{a}{R} \cdot \left( -1 + \sqrt{1 + \cfrac{2V}{\pi R a^2}}\right)}
      \end{equation}
Then the immersion distance $d_{sp/sp}$ must be found:
      \begin{equation}
        \label{eq:RAB:20}
        d_{sp/sp} = \cfrac{a}{2} \cdot \left[ -1 + \sqrt{1 + \cfrac{2V}{\pi R a^2}} \right]
      \end{equation}
Finally, $F_{cap}$ is predicted with:
      \begin{equation}
        \label{eq:RAB:19}
        F_{cap} = - \cfrac{2\pi R \gamma \cos{\theta}}{1 + \left[a / 2d_{sp/sp} \right]} - 2 \pi \gamma R \sin{\alpha} \sin{(\theta + \alpha)}
      \end{equation}
Later, Lambert et al. \cite{lambert2008comparison} identified an error in the 
deduction of the model and showed that the second term of Eq. (\ref{eq:RAB:19}) 
is redundant. Therefore, we employ Rabinovich's model with Lambert's correction. $F_{cap}$ 
is predicted with:
      \begin{equation}
        \label{eq:RABL:}
        F_{cap} = - \cfrac{2\pi R \gamma \cos{\theta}}{1 + \left[a / 2d_{sp/sp} \right]}
      \end{equation}
The sequence of equations in implementation is the following: Eq. (\ref{eq:RAB:A3}), 
Eq. (\ref{eq:RAB:20}), Eq. (\ref{eq:RABL:}).

\section{Validation of capillary bridge model implementations}
\label{exper}

Here, we observe outputs (forces) of different CBMs during the DEM pendular
simulation of a pair of particles with a single liquid bridge.
The setup of the simulations corresponds to the experiments that are presented in
Willett et al. \cite{Willett2000} for micro-scale and in Rabinovich et al. \cite{Rabinov2005}
for nano-scale experiments. 

Willett et al. \cite{Willett2000} employed for the experiments precision synthetic  
sapphire spheres of radii 2.381, 1.588 and 1.191 mm. In their experiments, the liquid 
bridges are formed of dimethylsiloxane with relevant material parameters surface 
tension $\gamma=20.6$ mN/m and contact angle $\theta=0^\circ$. 
The liquid bridge volume $V$ has been varied, see Tab. (\ref{tab:DEMSetup}). 
  
Rabinovich et al. \cite{Rabinov2005} used in their study much smaller glass 
spheres with the radii $R = [19 \ldots 35] \mu$m. An oil with relevant material parameters 
$\gamma$=$[24 \ldots 28]$mN/m  and $\theta=[0\ldots 10]^\circ$ forms the capillary bridges between the 
particles. All parameters for DEM-simulations are presented in Tab. (\ref{tab:DEMSetup}).
  
  \begin{table}
  \caption{Input values for DEM-simulations of a single pair of particles with a
          liquid bridge between them, based on Willett's \cite{Willett2000} and
          Rabinovich's \cite{Rabinov2005} experiments}
  \label{tab:DEMSetup} 
    \begin{tabular}{llllll}
      \hline\noalign{\smallskip}
      CODE & $R_i$ & $R_j$ & $\gamma$  & $\theta$       & $V$   \\
           & {[}mm]& [mm]  & [mN/m]    &[$\degree$]     & [nl]  \\
      \noalign{\smallskip}\hline\noalign{\smallskip}
      W$_{11}$ & 2.381 & 2.381 & 20.6      & 0          & 13.6  \\
      W$_{12}$ & 2.381 & 2.381 & 20.6      & 0          & 31.3  \\
      W$_{13}$ & 2.381 & 2.381 & 20.6      & 0          & 74.2  \\
      W$_{21}$ & 2.381 & 1.588 & 20.6      & 0          & 9.6   \\
      W$_{22}$ & 2.381 & 1.588 & 20.6      & 0          & 13.2  \\
      W$_{23}$ & 2.381 & 1.588 & 20.6      & 0          & 24.7  \\
      W$_{24}$ & 2.381 & 1.588 & 20.6      & 0          & 59.3  \\
      W$_{31}$ & 2.381 & 1.191 & 20.6      & 0          & 25.3  \\
      W$_{32}$ & 2.381 & 1.191 & 20.6      & 0          & 61.8  \\
      W$_{33}$ & 2.381 & 1.191 & 20.6      & 0          & 127.8 \\
      \noalign{\smallskip}\hline\noalign{\smallskip}
      & {[}$\mu$m]  & [$\mu$m]  & [mN/m]    & [$\degree$]   & [$\times10^8$nm$^3$]\\
      \noalign{\smallskip}\hline\noalign{\smallskip}
      R$_{1}$ &19    & 35    & 27   & 10 & 2  \\
      R$_{2}$ &19    & 32.5  & 24   & 10 & 12 \\
      R$_{3}$ &19    & 27.5  & 28   & 10 & 36 \\
      \noalign{\smallskip}\hline
    \end{tabular}
  \end{table}
  
Each of the 13 cases given in Tab. (\ref{tab:DEMSetup}) was investigated in DEM 
simulations using the four CBMs of Weigert et al. (Weig), Willett et al. - full (WilF), 
Willett et al. - reduced (WilR) and Rabinovich et al. with Lambert's correction 
(RabL).
  
  \begin{figure}[h!]
    \begin{center}
      \input{willett_1a_1.tex} 
    \end{center}
    \caption{Capillary force as a function of separation distance, comparison of
             Willett experiments with simulations, \cite[Fig. 1a, S. 9399]{Willett2000},
             W$_{11}$, see Tab. (\ref{tab:DEMSetup}): $R_{i}=R_{j}=2.381$ mm, $V=13.6$ nl}
    \label{fig:willett_1a_1}
  \end{figure}
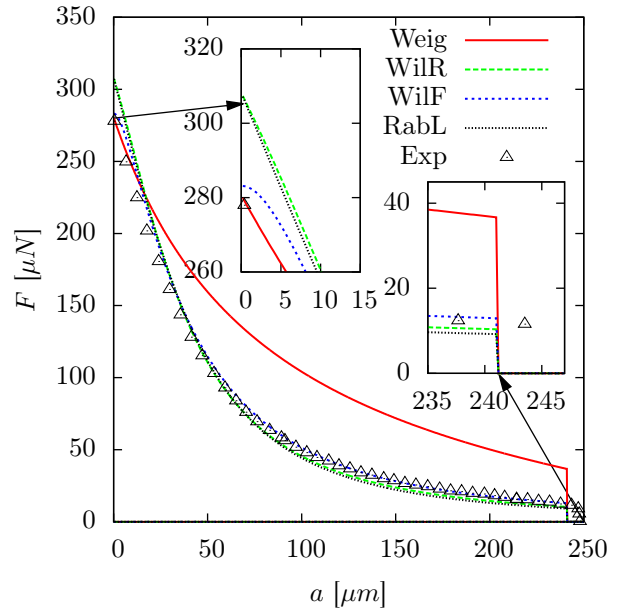

   \begin{figure}[h!]
    \begin{center}
      \input{rabinovich_1.tex}
    \end{center}
    \caption{Capillary force as a function of separation distance, comparison of
             Rabinovich experiments with simulations, line 1, \cite[Fig. 4, S. 10996]{Rabinov2005},
             R$_{1}$, see Tab. (\ref{tab:DEMSetup}): $R_{i}=19 \mu$m,
             $R_{j}=35 \mu$m, $V=2\times10^8$ nm$^3$}
    \label{fig:rabinovich_1}
  \end{figure}
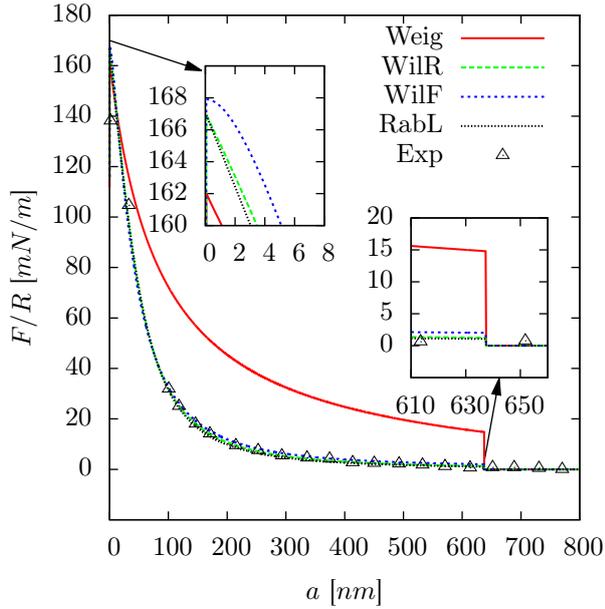
  
In the simulations, two interacting particles touch each other in their initial
positions without a gap between them, i. e. $a=0$. Then, one particle is pulled out
whereas the other remains fixed, until the
rupture distance of the liquid bridge $a_{crit}$ is reached. During the simulation, 
the force $F_{cap}$ of the capillary bridge is constantly recorded. Gravitation is 
not taken into account in the simulation.

In figures \ref{fig:willett_1a_1} and \ref{fig:rabinovich_1}, results for the 
capillary force calculation in two different DEM simulations (simulations W$_{11}$ and 
R$_1$ from Tab. (\ref{tab:DEMSetup}) are compared with the corresponding 
  experimental data. All simulations are performed until $a>a_{crit}$, therefore the jumps
  at the end of the curves represent the modeling of the liquid bridge rupture.

Obviously, the overall  agreement of the CBMs WilF, WilR and RabL to both 
experimental data sets from the micro- and the nano-scale is very good (see insets
on diagrams to get a zoomed view of the data). 
Differences between the CBMs predictions and measurements are largest for small 
particle distances $a$, i. e. small bridge elongations.
It looks like this difference is not systematic, but random.
Only in figure 2 Weig and WilF come close to experimental results for
$a \rightarrow 0$; in figure 3 all CBMs fail to predict the measured 
values for $F_{cap}$ in the case of $a \rightarrow 0$. We also have to
keep in mind that measurements of capillary forces at such small
separation distances are also complicated (see the discussion in \cite{Rabinov2005}).
  
On the contrary, the DEM simulations with CBM Weig fit only for small values of $a$ to 
the experimental findings. Especially for the micro-scale setup W$_{11}$ in figure 
\ref{fig:willett_1a_1}, the CBM Weig is noticeably better in 
comparison to the other CBMs. However, the prediction quality of the Weig CBM rapidly decreases 
for moderate and large values of $a$. Only the qualitative trend (nonlinear decreasing 
capillary force with increasing $a$) is captured with CBM Weig, but the capillary 
force values are markedly over-predicted.

  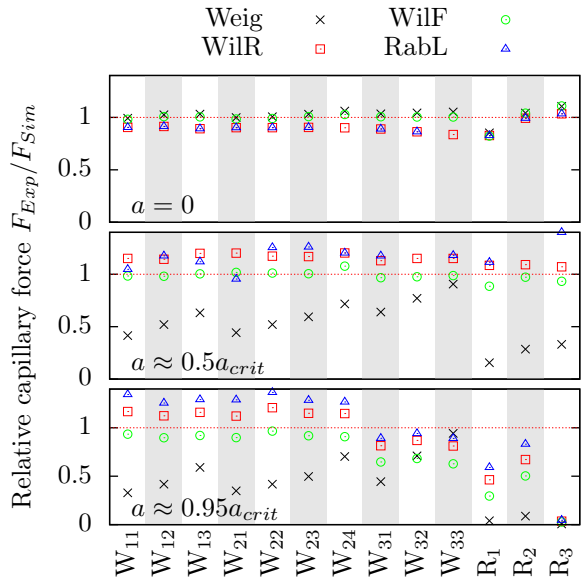
\begin{figure}[h!]
    \begin{center}
      \input{allPoints.tex}
    \end{center}
    \caption{Capillary forces, obtained from simulations, are compared with 
      experimental data (red dashed line)}
    \label{fig:allPoints}
  \end{figure}

An overview over the results from all DEM simulations (setups W$_{11}$ \dots W$_{33}$ 
and R$_1$ \dots R$_3$) is given in figure \ref{fig:allPoints}. Here, relative 
capillary force values, defined as the ratio of numerical to experimental force data, 
are given for three different bridge lengths $a_1=0$, $a_2=0.5\,a_{crit}$ and 
$a_3=0.95\,a_{crit}$. In general, the observations from setups W$_{11}$ and R$_{1}$ 
are confirmed: whereas CBMs WilF, WilR and RabL match well to the measurements for 
most setups and all particle distances, the Weig CBM only gives only a satisfactory prediction 
for small $a$ values.

\section{Influence on macroscopic parameters}
\label{macro}

\subsection{Shear cell setup}  

In a recent paper \cite{Schwarze2013a} the authors have demonstrated the visible
influence of capillary forces on macroscopic hydrodynamic material parameters, e.g. ``viscosity''
(ratio of shear stress to strain rate),
of wet granular matter in shearing motion. Now, the effects of the different CBMs on
these parameters are studied. Therefore, DEM-simulations of a split-bottom ring-shear
cell filled with a spherical beads are performed. The construction of such kind of
ring-shear cell was proposed and experimentally examined by Fenistein et al.
\cite{Fenistein2004}.
  
   \begin{figure}[h!]
    \begin{center}
       \includegraphics[width=1.00\linewidth]{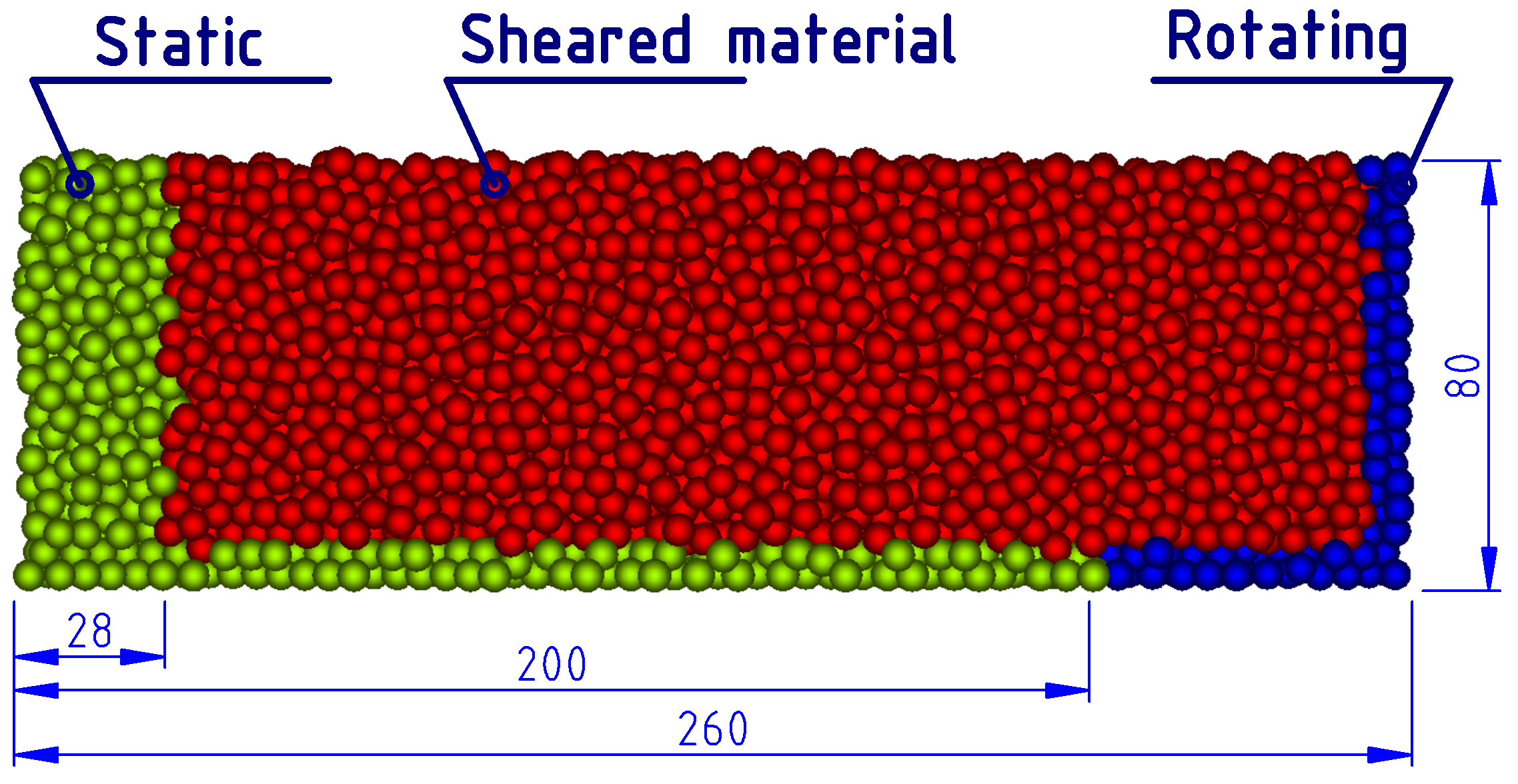}
    \end{center}
    \caption{Setup of the split-bottom ring-shear cell in the DEM simulations.
    The center line of the rotational symmetrical geometry is on the left, radii and 
    height are given in mm.}
    \label{fig:Schema}
  \end{figure}
  
The geometry of the packing in the shear cell is given in figure \ref{fig:Schema}. The 
particle size of the monodisperse packing has been set to $R_{i}=R_{j}=2.381$ mm, as in 
experiment $W_{11}$ in Tab. (\ref{tab:DEMSetup}). The outer cylinder and the outer part 
of the bottom  of the shear cell rotates with a period of $100$ s.  The inner cylinder 
and the inner part of the bottom remain static. 

The density of sheared particles is set to $\rho=150\,$ kg/m$^3$ in order to get 
roughly the same mass of particles as in \cite{Schwarze2013a} and achieve a
comparable $F_{norm}/F_{cap}$ relationship. The particle number for 
the simulation is nearly 200000. About 84\% of all particles are in the region of the 
sheared material. The following model parameters are set: duration of contact $t_{c} =
5.4\cdot 10^{-4}$ s, coefficient of restitution in normal and tangential directions 
$e_{n} = e_{t} = 0.83$, integration time step $\Delta t = 5.4 \cdot 10^{-4}\,$s.
According to formulation in \cite{Pournin2001} spring and 
damping coefficient under these conditions are the following: $k_{n} \approx 144.03 $ N/m,
$k_{t} \approx 41.15$ N/m, $c_{n} \approx 2.92 \cdot 10^{-3}$ kg/s and
$c_{t} \approx 41.15$ kg/s. The Coulomb friction of the contact is set to $\mu=0$ 
to exclude the influence of this parameter on the final results.

These settings of material parameters for the shear cell setup are purely
artificial and do not correspond to any experiments.
They have been chosen to see the effects of capillary bridges
more clearly. Simulations with more realistic material parameters
and comparison with data from shear cell experiments (not available
at the moment) are planned for the future.
 
Liquid bridge volume values for simulations of weakly wetted material have been chosen to be $13.6$ and
$74.2$\,nl, which correspond to experiments of single pairs of particles $W_{11}$ and $W_{13}$
from Tab. (\ref{tab:DEMSetup}) respectively. The distribution of capillary contacts is homogeneous 
without liquid conservation for the whole system. There are no interrelations between
individual capillary bridges. Therefore, a slight fluctuation of up 
to $1\,\%$ in the total liquid content was observed in the simulations.

Every simulation has been run for a minimum of $5$ s real flow time, corresponding to 
$\approx 10^6$ simulation time steps. 
The potential and kinetic energies of all simulated particles were controlled.
The results showed that such a low relatively simulation time is enough to 
achieve a steady state of the velocity field. This is in agreement with  
experimental studies of ring shear cell configurations \cite{Fenistein2004,fenistein2006core,Sakaie2008}.
However the whole system may also show some long-term transition in other quantities, e.g.
density, which was recently shown by Sakaie et al. \cite{Sakaie2008}. 
We do not consider these long-term developments, because we are focusing here on the
velocity field analysis. The influence of the long-term behavior will be
investigated for further studies.

All simulations have been carried out on the HPC cluster 
of TU Bergakademie Freiberg. Adding capillary bridges to the contact model of 
the DEM code increases the time of contact existence and the number of interactions.
It complicates the calculation schema, but does not significantly decrease the calculation speed of the model. All 
simulations need nearly the same CPU time period with differences of only $\approx 10 
\%$. From every simulation, 15 snapshots are taken into account for averaging, see 
\cite{Schwarze2013a} for more details. The discrete averaging time step is $0.25\,$s. 

\subsection{Results and discussion}
\label{compar}
  
For the analysis, data obtained from DEM simulations was averaged by micro-macro
transition as explained e.g. in \cite{luding2008,Schwarze2013a}. Therefore,
all quantities, e.g. velocity, stress and strain presented in the section
are macro-quantities.
  
\begin{figure}[h!]
    \begin{center}
      \input{Scherband.tex}
    \end{center}
    \caption{Shear bands and rates in dry (black thick line) and wet  (dashed thin lines) materials. Lines indicate the
             centers $R_{C}$ and widths $W$ of shear bands, which were obtained through
             the fit of error function Eq. (\ref{eq:errF}). Horizontal red lines on the lower plot are the heights at which the
             velocity profiles were analyzed.}
    \label{fig:Scherband}
  \end{figure}
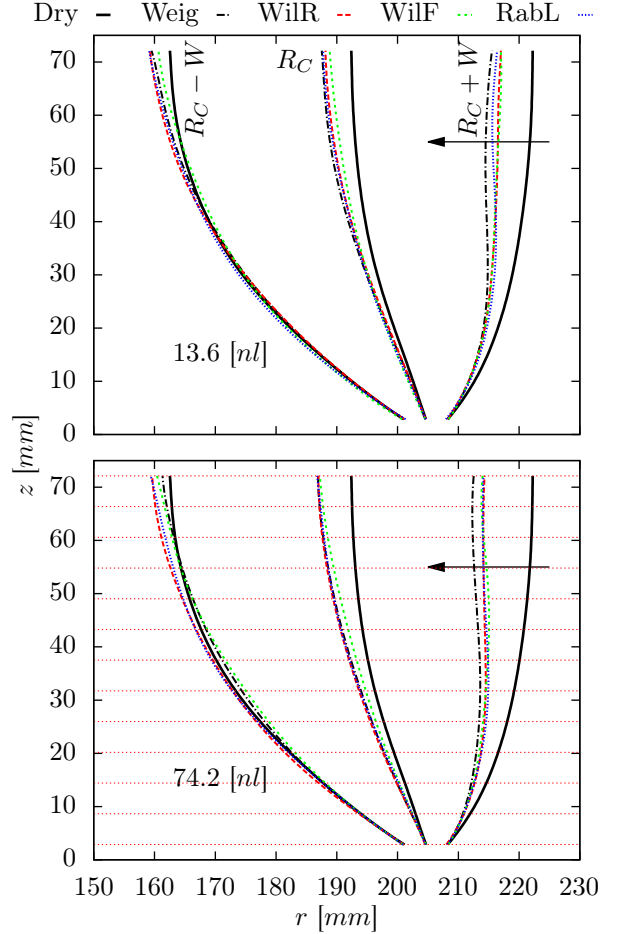

The dominating feature in the sheared granular material in the split-bottom shear cell 
is the shear band, i.e. a localized region where the granular material yields and 
flows \cite{Fenistein2004}. In figure \ref{fig:Scherband}, the shear bands in dry 
and wet granular materials are compared. For the wet materials, resulting shear 
bands from the four CBMs Weig, WilF, WilR and RabL are shown. 

Shear bands are indicated in figure \ref{fig:Scherband} by lines. They were 
obtained through the fit of error function Eq. (\ref{eq:errF}), which describes the universal 
velocity profile (see for details \cite{Fenistein2004,fenistein2006core,luding2008,Schwarze2013a})
of the sheared material. The central points of the shear band and its width were 
obtained for 13 different heights of the sheared material
  \begin{equation}
    \label{eq:errF}
      \omega(r) = 0.5 + 0.5 \cdot erf \left( \cfrac{r - {R_{C}}}{W} \right)
  \end{equation}
where $\omega(r)$ is the velocity profile at the defined height of the layer, 
$R_{C}$ and $W$ are shear band center and width respectively.

Obviously, the center $R_{C}$ and outer lines $R_{C}+W$ of the shear band are shifted inwards 
in the wet materials. The displacement seems to depend on the liquid content as well.
For the larger liquid bridge volume (higher liquid content in the granular material),
a stronger shift is observed. Regarding the four employed CBMs, we do not find a
clear difference between the corresponding results (relocation of the shear bands).
Similar shear band shifts were also observed by Luding during investigation of the
friction influence on the sheared material \cite{luding2008}. 
  
  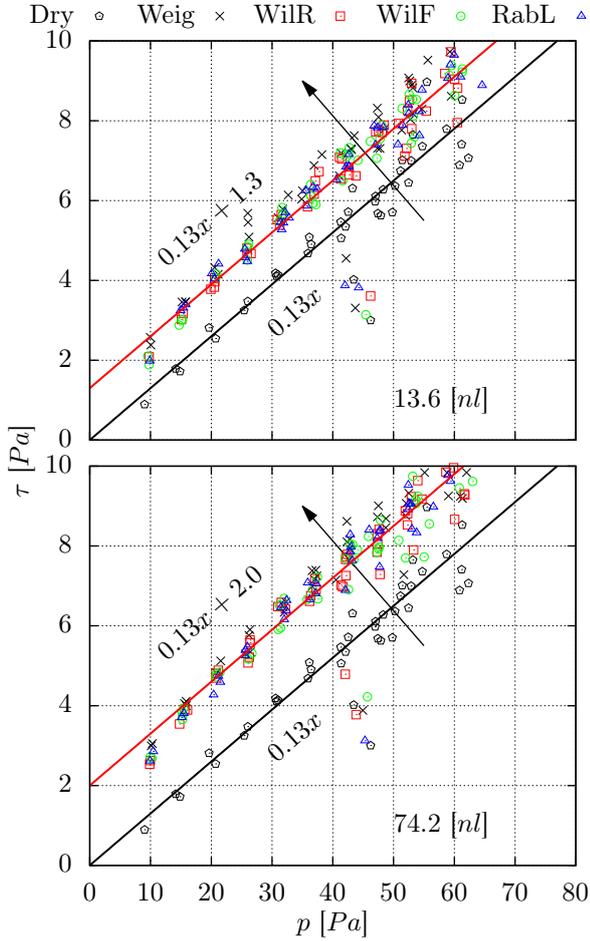
\begin{figure}[h!]
    \begin{center}
      \input{TauP.tex}
    \end{center}
    \caption{Shear stress $\tau$ plotted against pressure $p$ where $\dot{\gamma}>0.12$.
    Diagrams are showing the appeared cohesiveness of wet material.}
    \label{fig:TauP}
  \end{figure}

The mean shear and normal stresses correlations $\tau$-$p$ give first insights into 
the macroscopic behavior of the
granular material in the shear band. Figure \ref{fig:TauP} shows these correlations 
for dry and wet material with different liquid content. As expected, the correlations 
fit to the well-known constitutive law \cite{luding2011critical}
\begin{equation}
  \label{eq:mup}
  \tau = \mu\,p + c
\end{equation}
with $\mu = 0.13$ for the particular granular material. Here, the parameter $c$ of the
constitutive law depends significantly on the liquid content ($c_1 = 1.3\,Pa$ for 
$V_1 = 13.6$ nl, $c_2 = 2\,Pa$ for $V_2 = 74.2$ nl), whereas the 
choice of the CBM has again no obvious influence on the macroscopic parameters
(all points fitted by the red lines are very close to each other). It is definitely seen
that the liquid bridge existence and liquid content amount is increasing cohesiveness
of the simulated media. The increase of the parameter $c$ is due to the larger
rupture distance $a_{crit}$, which is increasing proportionally to the
liquid bridge volume. Therefore, the capillary forces act longer in the case of larger
liquid bridge volumes.
However the macro-correlations cannot be easily anticipated from the functional 
relationship given by the CBM.
We have researched for such simple correlations between micro- and macro-scale
functional relationships, but we have not found any at the moment.

The influence of the Coulomb friction parameter $\mu_{C}$ on shear band
structure and parameter $\mu$ is the same as reported by Luding \cite{luding2008}
for dry materials. Therefore a discussion is omitted here.

  \begin{figure}[h!]
    \begin{center}
      \input{torque_vergl.tex}
    \end{center}
    \caption{Average torque acting on rotating part of the shear cell during different simulation regimes.}
    \label{fig:torque_vergl}
  \end{figure}
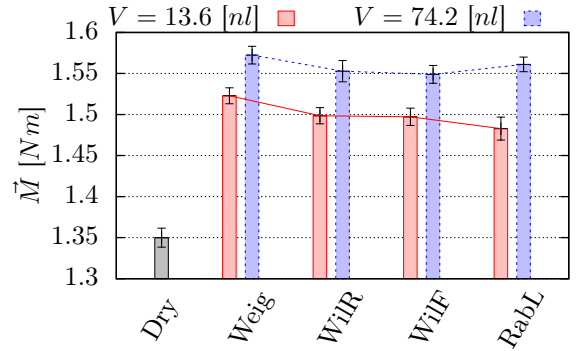
  
One more macro-parameter, which can be evaluated from the DEM results, is the torque $M$ 
acting on the rotating part of the shear cell. The torque indicates indirectly the cohesive properties 
of the sheared material as well. It has been recorded during the whole rotating period of 
the simulations. Time-averaged results are presented in figure \ref{fig:torque_vergl}. 
The results show the nonlinear dependency of the torque on the liquid content in the 
granular material, which can be the result of nonlinear CBM dependency. Again, the DEM simulations with the four different CBMs give nearly 
the same values of $M$.

We conclude from the results in figures \ref{fig:Scherband} to \ref{fig:torque_vergl}, 
that the specific choice of the CBM has a minor importance for this type of granular 
flow. Changes in the overall flow field structure and corresponding hydrodynamic 
parameters correlate clearly with the liquid content in the granular material, but 
only slightly with details of the capillary bridge force modelling.

  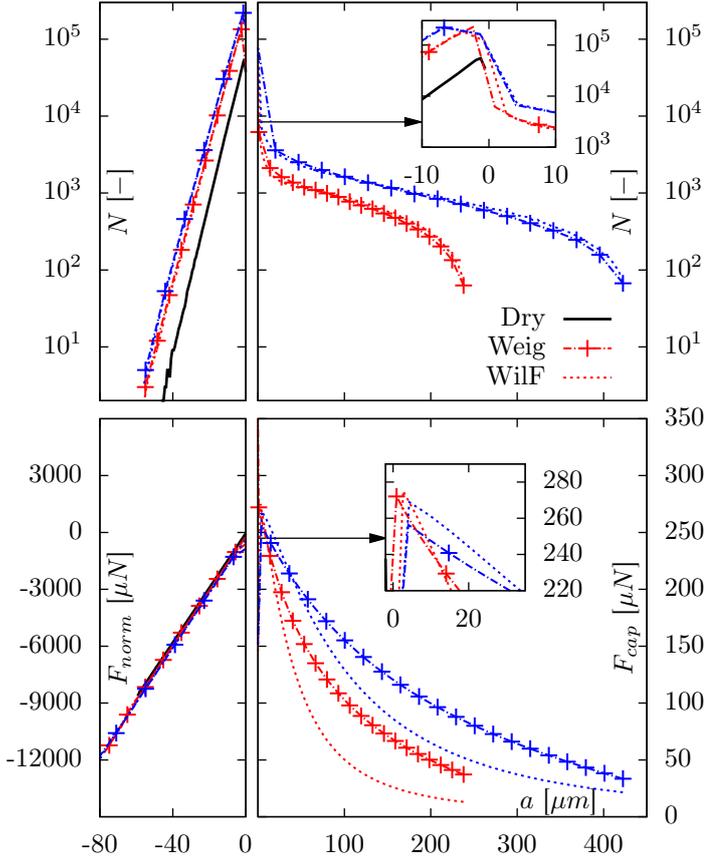
\begin{figure}[h!]
    \begin{center}
      \input{contactsA.tex}
    \end{center}
    \caption{Contact number as a function of separation distance $a$ (top row)
             for CBMs Weig and WilF. The bottom row shows the averaged force as a
             function of separation distance (right side) / penetration (left side)
             in a bulk. Blue lines -- $74.2$ [$nl$], red lines -- $13.6$ [$nl$].
             (WilR and RabL are excluded from the diagram, it is difficult to distinguish them from WilF).
             }
    \label{fig:ContactVert}
  \end{figure}

In order to understand these findings, we analyze the liquid bridge length 
distribution within the shear zone in more detail. Figure \ref{fig:ContactVert} 
gives the contact number and force distribution in both dry (mechanical contact $a=-\delta_{n}$) and wet (bridge length $a>0$)
contacts respectively. The averaging is performed as given in Eq. (\ref{avNF})
\begin{equation}
  \Phi(a) = \dfrac{1}{\Delta a} \int \limits_{a-\Delta a/2}^{a+\Delta a/2}\,\phi\left(a\right)\,da
  \label{avNF}
\end{equation}
for the contact number $N$ and the force $F$, respectively. The averaging interval is 
$\Delta a = \left(a_{crit}+ \delta_{n,max}\right)/100$, 
where  $\delta_{n,max}$ is the maximum penetration distance of spheres found in
the simulation. The inlay in the figure \ref{fig:ContactVert} (bottom row) shows,
that the maximum of capillary force values is not exactly at $a=0$ as can be seen in
figure \ref{fig:willett_1a_1}. That is an artifact due to the averaging interval, which
is not present in simulation of an individual particle pair.

It is found that in the simulation of weakly wetted granular material, the number
of mechanical contacts increases by about $20\%$, and is nearly independent of the individual 
liquid bridge volume. Inspecting the wet contacts, one can observe that most of them have only short 
bridge lengths with respect to $a_{crit}$. Here, the capillary force prediction gives 
roughly the same values from all four investigated CBMs. Therefore, only small 
deviations are found for the macroscopic flow parameters.

Also, the figure \ref{fig:ContactVert} (bottom row) shows distance-force dependencies, 
obtained from the shear cell simulations. Those curves have
the same dependency as in the previous sphere-sphere simulations (figure 
\ref{fig:willett_1a_1}). It proves the correct implementation of CBMs in the source
code. Under the more dynamic conditions (increased cell rotation speed
and thereby higher impact velocities), obtained curves can differ in a dry state 
from static ones because of the increased damping term of the total contact force.

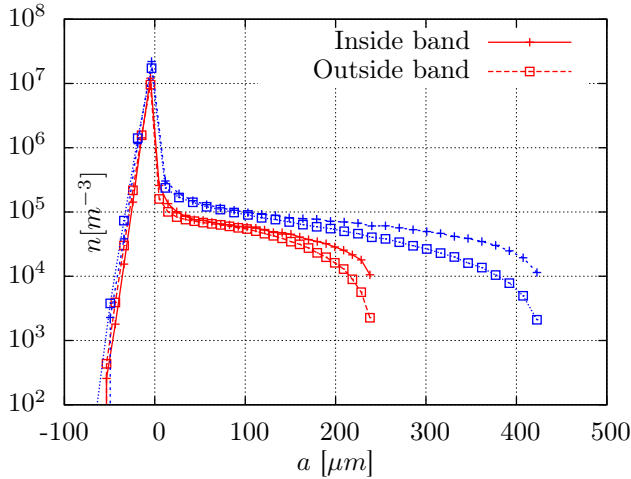
\begin{figure}[h!]
  \begin{center}
    \input{ContactVert.tex}
  \end{center}
  \caption{Contact density as a function of separation distance $a$
  inside shear band and outside of it for RabL CBM. 
  Blue symbols -- $74.2$ [$nl$],  red lines -- $13.6$ [$nl$]}
  \label{fig:ContactVertBand}
\end{figure}

Figure \ref{fig:ContactVertBand} shows the typical picture of wet contact 
distribution inside of the shear band and out of it. To correctly compare contact
numbers, the wet contact number per volume $n$ is introduced here. The diagram shows that
the relative number of contacts varies only for the range $a>0.5 a_{crit}$. 

The relative contact number is smaller for the range  $a>0.5 a_{crit}$ outside 
the shear band compared to the region of the shear band itself. The trend is 
independent of liquid volume. In other words, in the shear band there are some
more wet contacts, which are far away from each other due to the higher velocity
gradient.
 
\section{Conclusion}

\label{conclus}

In the paper, four different capillary bridge models for DEM 
simulations are investigated. First, implementations
of the four capillary bridge models in DEM are validated by a comparison of simulation
results and experimental measurements of an individual micro-scale pendular bridge between
two separating spheres.
It is found that the results of three capillary bridge models are very near to 
the  experimental data. In contrast, one model only gives a reasonable qualitative 
representation of the capillary force development, but overestimates the force 
values strongly.

Then, the different capillary bridge models are employed in DEM simulations of
wet granular material sheared in a split-bottom shear cell. Here, the flow field
structure and corresponding macroscopic parameters clearly depend on the micro-scale  
parameter liquid bridge volume.
However, up to now, the investigated macroscopic rheological dependencies of the sheared material
cannot be easily anticipated from the functional microscopic relationships given by the CBM.

Additionally, the differences that have been found in the
prediction of individual bridges on the micro-scale are not visible on the macro-scale
in this specific configuration.
Therefore, the specific choice of the investigated capillary bridge model does not seem to have
a marked influence on the prediction of the hydrodynamics of this type of granular
flow. Separation distance $a$ for most of the wet contacts in the bulk is relatively
small in comparison to the critical distance ${a}_{crit}$, where all CBMs showed
comparable results. 

\section*{Disclosure}
The authors express their thanks to the Deutsche Forschungsgemeinschaft, which 
supported this work within the DFG/STW project SCHW 1168/6-1 ``Hydrodynamic theory of wet 
particle systems''.

\section*{Acknowledgments}
Special thanks go to Stefan Luding and our partners at TU Dortmund and UTwente 
in this project for the helpful discussion.

The final publication is available at \href{http://dx.doi.org/10.1007/s10035-014-0527-z}{www.springerlink.com}

%% file: willett_1a_1.tex
\begingroup
  \makeatletter
  \providecommand\color[2][]{%
    \GenericError{(gnuplot) \space\space\space\@spaces}{%
      Package color not loaded in conjunction with
      terminal option `colourtext'%
    }{See the gnuplot documentation for explanation.%
    }{Either use 'blacktext' in gnuplot or load the package
      color.sty in LaTeX.}%
    \renewcommand\color[2][]{}%
  }%
  \providecommand\includegraphics[2][]{%
    \GenericError{(gnuplot) \space\space\space\@spaces}{%
      Package graphicx or graphics not loaded%
    }{See the gnuplot documentation for explanation.%
    }{The gnuplot epslatex terminal needs graphicx.sty or graphics.sty.}%
    \renewcommand\includegraphics[2][]{}%
  }%
  \providecommand\rotatebox[2]{#2}%
  \@ifundefined{ifGPcolor}{%
    \newif\ifGPcolor
    \GPcolortrue
  }{}%
  \@ifundefined{ifGPblacktext}{%
    \newif\ifGPblacktext
    \GPblacktexttrue
  }{}%
  \let\gplgaddtomacro\g@addto@macro
  \gdef\gplbacktext{}%
  \gdef\gplfronttext{}%
  \makeatother
  \ifGPblacktext
    \def\colorrgb#1{}%
    \def\colorgray#1{}%
  \else
    \ifGPcolor
      \def\colorrgb#1{\color[rgb]{#1}}%
      \def\colorgray#1{\color[gray]{#1}}%
      \expandafter\def\csname LTw\endcsname{\color{white}}%
      \expandafter\def\csname LTb\endcsname{\color{black}}%
      \expandafter\def\csname LTa\endcsname{\color{black}}%
      \expandafter\def\csname LT0\endcsname{\color[rgb]{1,0,0}}%
      \expandafter\def\csname LT1\endcsname{\color[rgb]{0,1,0}}%
      \expandafter\def\csname LT2\endcsname{\color[rgb]{0,0,1}}%
      \expandafter\def\csname LT3\endcsname{\color[rgb]{1,0,1}}%
      \expandafter\def\csname LT4\endcsname{\color[rgb]{0,1,1}}%
      \expandafter\def\csname LT5\endcsname{\color[rgb]{1,1,0}}%
      \expandafter\def\csname LT6\endcsname{\color[rgb]{0,0,0}}%
      \expandafter\def\csname LT7\endcsname{\color[rgb]{1,0.3,0}}%
      \expandafter\def\csname LT8\endcsname{\color[rgb]{0.5,0.5,0.5}}%
    \else
      \def\colorrgb#1{\color{black}}%
      \def\colorgray#1{\color[gray]{#1}}%
      \expandafter\def\csname LTw\endcsname{\color{white}}%
      \expandafter\def\csname LTb\endcsname{\color{black}}%
      \expandafter\def\csname LTa\endcsname{\color{black}}%
      \expandafter\def\csname LT0\endcsname{\color{black}}%
      \expandafter\def\csname LT1\endcsname{\color{black}}%
      \expandafter\def\csname LT2\endcsname{\color{black}}%
      \expandafter\def\csname LT3\endcsname{\color{black}}%
      \expandafter\def\csname LT4\endcsname{\color{black}}%
      \expandafter\def\csname LT5\endcsname{\color{black}}%
      \expandafter\def\csname LT6\endcsname{\color{black}}%
      \expandafter\def\csname LT7\endcsname{\color{black}}%
      \expandafter\def\csname LT8\endcsname{\color{black}}%
    \fi
  \fi
  \setlength{\unitlength}{0.0500bp}%
  \begin{picture}(4762.00,4762.00)%
    \gplgaddtomacro\gplbacktext{%
      \csname LTb\endcsname%
      \put(726,704){\makebox(0,0)[r]{\strut{} 0}}%
      \put(726,1246){\makebox(0,0)[r]{\strut{} 50}}%
      \put(726,1788){\makebox(0,0)[r]{\strut{} 100}}%
      \put(726,2330){\makebox(0,0)[r]{\strut{} 150}}%
      \put(726,2871){\makebox(0,0)[r]{\strut{} 200}}%
      \put(726,3413){\makebox(0,0)[r]{\strut{} 250}}%
      \put(726,3955){\makebox(0,0)[r]{\strut{} 300}}%
      \put(726,4497){\makebox(0,0)[r]{\strut{} 350}}%
      \put(858,484){\makebox(0,0){\strut{} 0}}%
      \put(1559,484){\makebox(0,0){\strut{} 50}}%
      \put(2261,484){\makebox(0,0){\strut{} 100}}%
      \put(2962,484){\makebox(0,0){\strut{} 150}}%
      \put(3664,484){\makebox(0,0){\strut{} 200}}%
      \put(4365,484){\makebox(0,0){\strut{} 250}}%
      \put(220,2600){\rotatebox{-270}{\makebox(0,0){\strut{}$F$ [$\mu N$]}}}%
      \put(2611,154){\makebox(0,0){\strut{}$a$ [$\mu m$]}}%
    }%
    \gplgaddtomacro\gplfronttext{%
      \csname LTb\endcsname%
      \put(3347,4324){\makebox(0,0)[r]{\strut{}Weig}}%
      \csname LTb\endcsname%
      \put(3347,4104){\makebox(0,0)[r]{\strut{}WilR}}%
      \csname LTb\endcsname%
      \put(3347,3884){\makebox(0,0)[r]{\strut{}WilF}}%
      \csname LTb\endcsname%
      \put(3347,3664){\makebox(0,0)[r]{\strut{}RabL}}%
      \csname LTb\endcsname%
      \put(3347,3444){\makebox(0,0)[r]{\strut{}Exp}}%
    }%
    \gplgaddtomacro\gplbacktext{%
      \csname LTb\endcsname%
      \put(1678,2582){\makebox(0,0)[r]{\strut{} 260}}%
      \put(1678,3141){\makebox(0,0)[r]{\strut{} 280}}%
      \put(1678,3700){\makebox(0,0)[r]{\strut{} 300}}%
      \put(1678,4259){\makebox(0,0)[r]{\strut{} 320}}%
      \put(1810,2362){\makebox(0,0){\strut{} 0}}%
      \put(2106,2362){\makebox(0,0){\strut{} 5}}%
      \put(2402,2362){\makebox(0,0){\strut{} 10}}%
      \put(2698,2362){\makebox(0,0){\strut{} 15}}%
    }%
    \gplgaddtomacro\gplfronttext{%
    }%
    \gplgaddtomacro\gplbacktext{%
      \csname LTb\endcsname%
      \put(3070,1820){\makebox(0,0)[r]{\strut{} 0}}%
      \put(3070,2460){\makebox(0,0)[r]{\strut{} 20}}%
      \put(3070,3099){\makebox(0,0)[r]{\strut{} 40}}%
      \put(3202,1600){\makebox(0,0){\strut{} 235}}%
      \put(3627,1600){\makebox(0,0){\strut{} 240}}%
      \put(4052,1600){\makebox(0,0){\strut{} 245}}%
    }%
    \gplgaddtomacro\gplfronttext{%
    }%
    \gplbacktext
    \put(0,0){\includegraphics{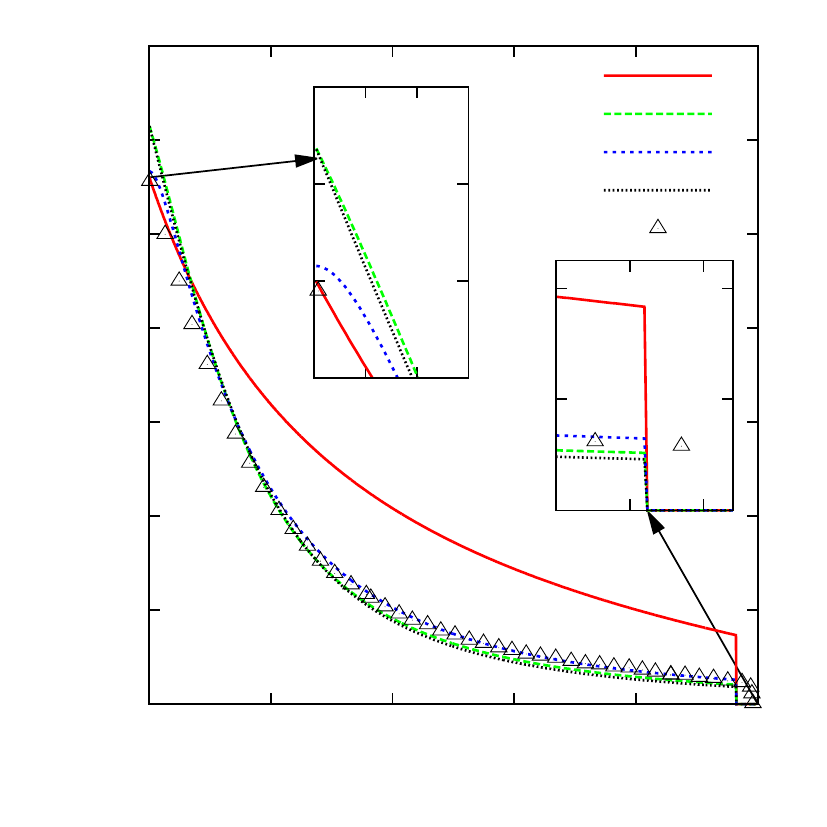}}%
    \gplfronttext
  \end{picture}%
\endgroup

%% file: rabinovich_1.tex
\begingroup
  \makeatletter
  \providecommand\color[2][]{%
    \GenericError{(gnuplot) \space\space\space\@spaces}{%
      Package color not loaded in conjunction with
      terminal option `colourtext'%
    }{See the gnuplot documentation for explanation.%
    }{Either use 'blacktext' in gnuplot or load the package
      color.sty in LaTeX.}%
    \renewcommand\color[2][]{}%
  }%
  \providecommand\includegraphics[2][]{%
    \GenericError{(gnuplot) \space\space\space\@spaces}{%
      Package graphicx or graphics not loaded%
    }{See the gnuplot documentation for explanation.%
    }{The gnuplot epslatex terminal needs graphicx.sty or graphics.sty.}%
    \renewcommand\includegraphics[2][]{}%
  }%
  \providecommand\rotatebox[2]{#2}%
  \@ifundefined{ifGPcolor}{%
    \newif\ifGPcolor
    \GPcolortrue
  }{}%
  \@ifundefined{ifGPblacktext}{%
    \newif\ifGPblacktext
    \GPblacktexttrue
  }{}%
  \let\gplgaddtomacro\g@addto@macro
  \gdef\gplbacktext{}%
  \gdef\gplfronttext{}%
  \makeatother
  \ifGPblacktext
    \def\colorrgb#1{}%
    \def\colorgray#1{}%
  \else
    \ifGPcolor
      \def\colorrgb#1{\color[rgb]{#1}}%
      \def\colorgray#1{\color[gray]{#1}}%
      \expandafter\def\csname LTw\endcsname{\color{white}}%
      \expandafter\def\csname LTb\endcsname{\color{black}}%
      \expandafter\def\csname LTa\endcsname{\color{black}}%
      \expandafter\def\csname LT0\endcsname{\color[rgb]{1,0,0}}%
      \expandafter\def\csname LT1\endcsname{\color[rgb]{0,1,0}}%
      \expandafter\def\csname LT2\endcsname{\color[rgb]{0,0,1}}%
      \expandafter\def\csname LT3\endcsname{\color[rgb]{1,0,1}}%
      \expandafter\def\csname LT4\endcsname{\color[rgb]{0,1,1}}%
      \expandafter\def\csname LT5\endcsname{\color[rgb]{1,1,0}}%
      \expandafter\def\csname LT6\endcsname{\color[rgb]{0,0,0}}%
      \expandafter\def\csname LT7\endcsname{\color[rgb]{1,0.3,0}}%
      \expandafter\def\csname LT8\endcsname{\color[rgb]{0.5,0.5,0.5}}%
    \else
      \def\colorrgb#1{\color{black}}%
      \def\colorgray#1{\color[gray]{#1}}%
      \expandafter\def\csname LTw\endcsname{\color{white}}%
      \expandafter\def\csname LTb\endcsname{\color{black}}%
      \expandafter\def\csname LTa\endcsname{\color{black}}%
      \expandafter\def\csname LT0\endcsname{\color{black}}%
      \expandafter\def\csname LT1\endcsname{\color{black}}%
      \expandafter\def\csname LT2\endcsname{\color{black}}%
      \expandafter\def\csname LT3\endcsname{\color{black}}%
      \expandafter\def\csname LT4\endcsname{\color{black}}%
      \expandafter\def\csname LT5\endcsname{\color{black}}%
      \expandafter\def\csname LT6\endcsname{\color{black}}%
      \expandafter\def\csname LT7\endcsname{\color{black}}%
      \expandafter\def\csname LT8\endcsname{\color{black}}%
    \fi
  \fi
  \setlength{\unitlength}{0.0500bp}%
  \begin{picture}(4762.00,4762.00)%
    \gplgaddtomacro\gplbacktext{%
      \csname LTb\endcsname%
      \put(726,1083){\makebox(0,0)[r]{\strut{} 0}}%
      \put(726,1463){\makebox(0,0)[r]{\strut{} 20}}%
      \put(726,1842){\makebox(0,0)[r]{\strut{} 40}}%
      \put(726,2221){\makebox(0,0)[r]{\strut{} 60}}%
      \put(726,2601){\makebox(0,0)[r]{\strut{} 80}}%
      \put(726,2980){\makebox(0,0)[r]{\strut{} 100}}%
      \put(726,3359){\makebox(0,0)[r]{\strut{} 120}}%
      \put(726,3738){\makebox(0,0)[r]{\strut{} 140}}%
      \put(726,4118){\makebox(0,0)[r]{\strut{} 160}}%
      \put(726,4497){\makebox(0,0)[r]{\strut{} 180}}%
      \put(858,484){\makebox(0,0){\strut{} 0}}%
      \put(1296,484){\makebox(0,0){\strut{} 100}}%
      \put(1735,484){\makebox(0,0){\strut{} 200}}%
      \put(2173,484){\makebox(0,0){\strut{} 300}}%
      \put(2612,484){\makebox(0,0){\strut{} 400}}%
      \put(3050,484){\makebox(0,0){\strut{} 500}}%
      \put(3488,484){\makebox(0,0){\strut{} 600}}%
      \put(3927,484){\makebox(0,0){\strut{} 700}}%
      \put(4365,484){\makebox(0,0){\strut{} 800}}%
      \put(220,2600){\rotatebox{-270}{\makebox(0,0){\strut{}$F/R$ [$mN/m$]}}}%
      \put(2611,154){\makebox(0,0){\strut{}$a$ [$nm$]}}%
    }%
    \gplgaddtomacro\gplfronttext{%
      \csname LTb\endcsname%
      \put(3347,4324){\makebox(0,0)[r]{\strut{}Weig}}%
      \csname LTb\endcsname%
      \put(3347,4104){\makebox(0,0)[r]{\strut{}WilR}}%
      \csname LTb\endcsname%
      \put(3347,3884){\makebox(0,0)[r]{\strut{}WilF}}%
      \csname LTb\endcsname%
      \put(3347,3664){\makebox(0,0)[r]{\strut{}RabL}}%
      \csname LTb\endcsname%
      \put(3347,3444){\makebox(0,0)[r]{\strut{}Exp}}%
    }%
    \gplgaddtomacro\gplbacktext{%
      \csname LTb\endcsname%
      \put(1440,2916){\makebox(0,0)[r]{\strut{} 160}}%
      \put(1440,3156){\makebox(0,0)[r]{\strut{} 162}}%
      \put(1440,3396){\makebox(0,0)[r]{\strut{} 164}}%
      \put(1440,3636){\makebox(0,0)[r]{\strut{} 166}}%
      \put(1440,3876){\makebox(0,0)[r]{\strut{} 168}}%
      \put(1572,2696){\makebox(0,0){\strut{} 0}}%
      \put(1794,2696){\makebox(0,0){\strut{} 2}}%
      \put(2016,2696){\makebox(0,0){\strut{} 4}}%
      \put(2238,2696){\makebox(0,0){\strut{} 6}}%
      \put(2460,2696){\makebox(0,0){\strut{} 8}}%
    }%
    \gplgaddtomacro\gplfronttext{%
    }%
    \gplgaddtomacro\gplbacktext{%
      \csname LTb\endcsname%
      \put(2975,2013){\makebox(0,0)[r]{\strut{} 0}}%
      \put(2975,2253){\makebox(0,0)[r]{\strut{} 5}}%
      \put(2975,2493){\makebox(0,0)[r]{\strut{} 10}}%
      \put(2975,2733){\makebox(0,0)[r]{\strut{} 15}}%
      \put(2975,2973){\makebox(0,0)[r]{\strut{} 20}}%
      \put(3107,1553){\makebox(0,0){\strut{} 610}}%
      \put(3515,1553){\makebox(0,0){\strut{} 630}}%
      \put(3922,1553){\makebox(0,0){\strut{} 650}}%
    }%
    \gplgaddtomacro\gplfronttext{%
    }%
    \gplbacktext
    \put(0,0){\includegraphics{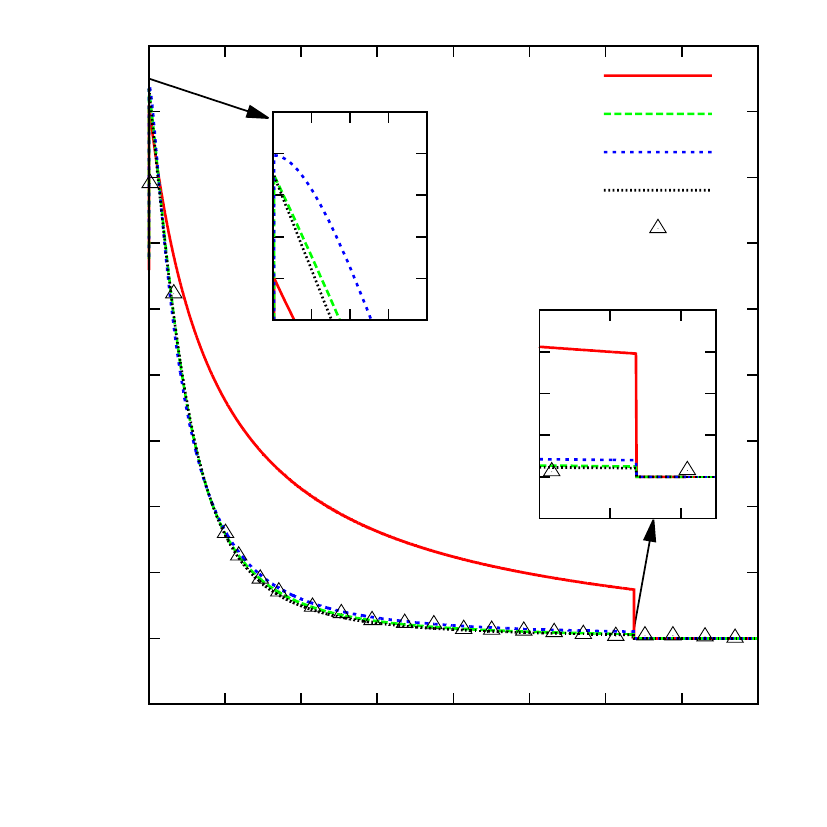}}%
    \gplfronttext
  \end{picture}%
\endgroup

%% file: allPoints.tex
\begingroup
  \makeatletter
  \providecommand\color[2][]{%
    \GenericError{(gnuplot) \space\space\space\@spaces}{%
      Package color not loaded in conjunction with
      terminal option `colourtext'%
    }{See the gnuplot documentation for explanation.%
    }{Either use 'blacktext' in gnuplot or load the package
      color.sty in LaTeX.}%
    \renewcommand\color[2][]{}%
  }%
  \providecommand\includegraphics[2][]{%
    \GenericError{(gnuplot) \space\space\space\@spaces}{%
      Package graphicx or graphics not loaded%
    }{See the gnuplot documentation for explanation.%
    }{The gnuplot epslatex terminal needs graphicx.sty or graphics.sty.}%
    \renewcommand\includegraphics[2][]{}%
  }%
  \providecommand\rotatebox[2]{#2}%
  \@ifundefined{ifGPcolor}{%
    \newif\ifGPcolor
    \GPcolortrue
  }{}%
  \@ifundefined{ifGPblacktext}{%
    \newif\ifGPblacktext
    \GPblacktexttrue
  }{}%
  \let\gplgaddtomacro\g@addto@macro
  \gdef\gplbacktext{}%
  \gdef\gplfronttext{}%
  \makeatother
  \ifGPblacktext
    \def\colorrgb#1{}%
    \def\colorgray#1{}%
  \else
    \ifGPcolor
      \def\colorrgb#1{\color[rgb]{#1}}%
      \def\colorgray#1{\color[gray]{#1}}%
      \expandafter\def\csname LTw\endcsname{\color{white}}%
      \expandafter\def\csname LTb\endcsname{\color{black}}%
      \expandafter\def\csname LTa\endcsname{\color{black}}%
      \expandafter\def\csname LT0\endcsname{\color[rgb]{1,0,0}}%
      \expandafter\def\csname LT1\endcsname{\color[rgb]{0,1,0}}%
      \expandafter\def\csname LT2\endcsname{\color[rgb]{0,0,1}}%
      \expandafter\def\csname LT3\endcsname{\color[rgb]{1,0,1}}%
      \expandafter\def\csname LT4\endcsname{\color[rgb]{0,1,1}}%
      \expandafter\def\csname LT5\endcsname{\color[rgb]{1,1,0}}%
      \expandafter\def\csname LT6\endcsname{\color[rgb]{0,0,0}}%
      \expandafter\def\csname LT7\endcsname{\color[rgb]{1,0.3,0}}%
      \expandafter\def\csname LT8\endcsname{\color[rgb]{0.5,0.5,0.5}}%
    \else
      \def\colorrgb#1{\color{black}}%
      \def\colorgray#1{\color[gray]{#1}}%
      \expandafter\def\csname LTw\endcsname{\color{white}}%
      \expandafter\def\csname LTb\endcsname{\color{black}}%
      \expandafter\def\csname LTa\endcsname{\color{black}}%
      \expandafter\def\csname LT0\endcsname{\color{black}}%
      \expandafter\def\csname LT1\endcsname{\color{black}}%
      \expandafter\def\csname LT2\endcsname{\color{black}}%
      \expandafter\def\csname LT3\endcsname{\color{black}}%
      \expandafter\def\csname LT4\endcsname{\color{black}}%
      \expandafter\def\csname LT5\endcsname{\color{black}}%
      \expandafter\def\csname LT6\endcsname{\color{black}}%
      \expandafter\def\csname LT7\endcsname{\color{black}}%
      \expandafter\def\csname LT8\endcsname{\color{black}}%
    \fi
  \fi
  \setlength{\unitlength}{0.0500bp}%
  \begin{picture}(4762.00,4534.00)%
    \gplgaddtomacro\gplbacktext{%
      \csname LTb\endcsname%
      \put(726,2713){\makebox(0,0)[r]{\strut{} 0}}%
      \put(726,3107){\makebox(0,0)[r]{\strut{} 0.5}}%
      \put(726,3501){\makebox(0,0)[r]{\strut{} 1}}%
      \put(993,2317){\rotatebox{90}{\makebox(0,0)[l]{\strut{}}}}%
      \put(1532,2317){\rotatebox{90}{\makebox(0,0)[l]{\strut{}}}}%
      \put(2072,2317){\rotatebox{90}{\makebox(0,0)[l]{\strut{}}}}%
      \put(2612,2317){\rotatebox{90}{\makebox(0,0)[l]{\strut{}}}}%
      \put(3151,2317){\rotatebox{90}{\makebox(0,0)[l]{\strut{}}}}%
      \put(3691,2317){\rotatebox{90}{\makebox(0,0)[l]{\strut{}}}}%
      \put(4230,2317){\rotatebox{90}{\makebox(0,0)[l]{\strut{}}}}%
      \put(352,3264){\rotatebox{-270}{\makebox(0,0){\strut{} }}}%
    }%
    \gplgaddtomacro\gplfronttext{%
      \csname LTb\endcsname%
      \put(2002,4235){\makebox(0,0)[r]{\strut{}Weig}}%
      \csname LTb\endcsname%
      \put(2002,4015){\makebox(0,0)[r]{\strut{}WilR}}%
      \csname LTb\endcsname%
      \put(3385,4235){\makebox(0,0)[r]{\strut{}WilF}}%
      \csname LTb\endcsname%
      \put(3385,4015){\makebox(0,0)[r]{\strut{}RabL}}%
      \csname LTb\endcsname%
      \put(1020,2831){\makebox(0,0)[l]{\strut{}$a=0$}}%
    }%
    \gplgaddtomacro\gplbacktext{%
      \csname LTb\endcsname%
      \put(726,1534){\makebox(0,0)[r]{\strut{} 0}}%
      \put(726,1928){\makebox(0,0)[r]{\strut{} 0.5}}%
      \put(726,2322){\makebox(0,0)[r]{\strut{} 1}}%
      \put(993,1138){\rotatebox{90}{\makebox(0,0)[l]{\strut{}}}}%
      \put(1532,1138){\rotatebox{90}{\makebox(0,0)[l]{\strut{}}}}%
      \put(2072,1138){\rotatebox{90}{\makebox(0,0)[l]{\strut{}}}}%
      \put(2612,1138){\rotatebox{90}{\makebox(0,0)[l]{\strut{}}}}%
      \put(3151,1138){\rotatebox{90}{\makebox(0,0)[l]{\strut{}}}}%
      \put(3691,1138){\rotatebox{90}{\makebox(0,0)[l]{\strut{}}}}%
      \put(4230,1138){\rotatebox{90}{\makebox(0,0)[l]{\strut{}}}}%
      \put(220,2085){\rotatebox{-270}{\makebox(0,0){\strut{}Relative capillary force $F_{Exp}/F_{Sim}$}}}%
    }%
    \gplgaddtomacro\gplfronttext{%
      \csname LTb\endcsname%
      \put(1020,1652){\makebox(0,0)[l]{\strut{}$a \approx 0.5 a_{crit}$}}%
    }%
    \gplgaddtomacro\gplbacktext{%
      \csname LTb\endcsname%
      \put(726,440){\makebox(0,0)[r]{\strut{} 0}}%
      \put(726,804){\makebox(0,0)[r]{\strut{} 0.5}}%
      \put(726,1168){\makebox(0,0)[r]{\strut{} 1}}%
      \put(993,44){\rotatebox{90}{\makebox(0,0)[l]{\strut{}W$_{11}$}}}%
      \put(1263,44){\rotatebox{90}{\makebox(0,0)[l]{\strut{}W$_{12}$}}}%
      \put(1532,44){\rotatebox{90}{\makebox(0,0)[l]{\strut{}W$_{13}$}}}%
      \put(1802,44){\rotatebox{90}{\makebox(0,0)[l]{\strut{}W$_{21}$}}}%
      \put(2072,44){\rotatebox{90}{\makebox(0,0)[l]{\strut{}W$_{22}$}}}%
      \put(2342,44){\rotatebox{90}{\makebox(0,0)[l]{\strut{}W$_{23}$}}}%
      \put(2612,44){\rotatebox{90}{\makebox(0,0)[l]{\strut{}W$_{24}$}}}%
      \put(2881,44){\rotatebox{90}{\makebox(0,0)[l]{\strut{}W$_{31}$}}}%
      \put(3151,44){\rotatebox{90}{\makebox(0,0)[l]{\strut{}W$_{32}$}}}%
      \put(3421,44){\rotatebox{90}{\makebox(0,0)[l]{\strut{}W$_{33}$}}}%
      \put(3691,44){\rotatebox{90}{\makebox(0,0)[l]{\strut{}R$_{1}$}}}%
      \put(3960,44){\rotatebox{90}{\makebox(0,0)[l]{\strut{}R$_{2}$}}}%
      \put(4230,44){\rotatebox{90}{\makebox(0,0)[l]{\strut{}R$_{3}$}}}%
      \put(352,949){\rotatebox{-270}{\makebox(0,0){\strut{} }}}%
    }%
    \gplgaddtomacro\gplfronttext{%
      \csname LTb\endcsname%
      \put(1020,549){\makebox(0,0)[l]{\strut{}$a\approx 0.95 a_{crit}$}}%
    }%
    \gplbacktext
    \put(0,0){\includegraphics{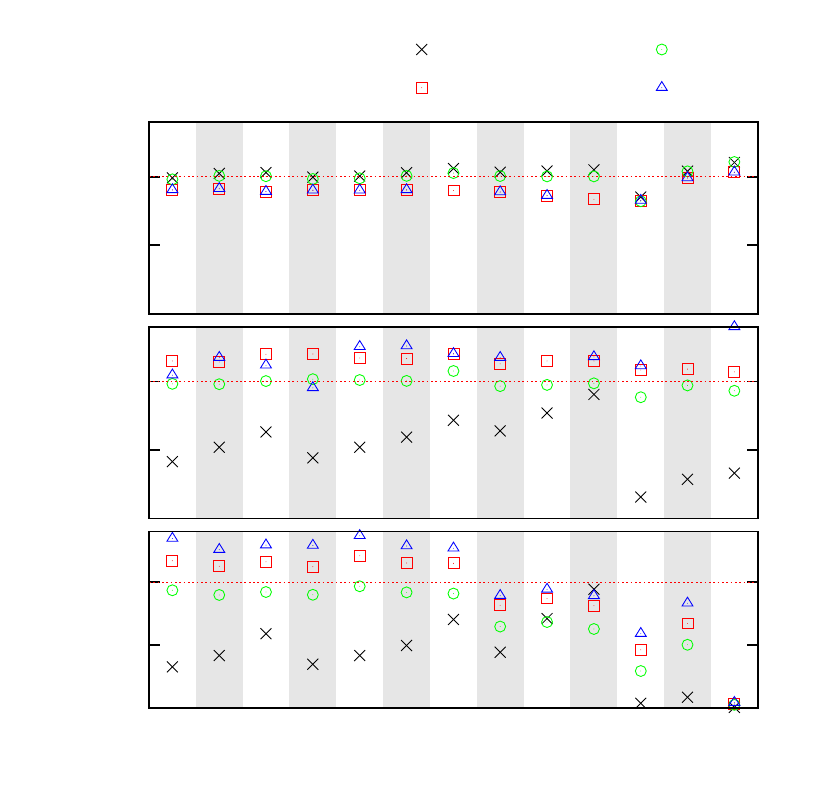}}%
    \gplfronttext
  \end{picture}%
\endgroup

%% file: Scherband.tex
\begingroup
  \makeatletter
  \providecommand\color[2][]{%
    \GenericError{(gnuplot) \space\space\space\@spaces}{%
      Package color not loaded in conjunction with
      terminal option `colourtext'%
    }{See the gnuplot documentation for explanation.%
    }{Either use 'blacktext' in gnuplot or load the package
      color.sty in LaTeX.}%
    \renewcommand\color[2][]{}%
  }%
  \providecommand\includegraphics[2][]{%
    \GenericError{(gnuplot) \space\space\space\@spaces}{%
      Package graphicx or graphics not loaded%
    }{See the gnuplot documentation for explanation.%
    }{The gnuplot epslatex terminal needs graphicx.sty or graphics.sty.}%
    \renewcommand\includegraphics[2][]{}%
  }%
  \providecommand\rotatebox[2]{#2}%
  \@ifundefined{ifGPcolor}{%
    \newif\ifGPcolor
    \GPcolortrue
  }{}%
  \@ifundefined{ifGPblacktext}{%
    \newif\ifGPblacktext
    \GPblacktexttrue
  }{}%
  \let\gplgaddtomacro\g@addto@macro
  \gdef\gplbacktext{}%
  \gdef\gplfronttext{}%
  \makeatother
  \ifGPblacktext
    \def\colorrgb#1{}%
    \def\colorgray#1{}%
  \else
    \ifGPcolor
      \def\colorrgb#1{\color[rgb]{#1}}%
      \def\colorgray#1{\color[gray]{#1}}%
      \expandafter\def\csname LTw\endcsname{\color{white}}%
      \expandafter\def\csname LTb\endcsname{\color{black}}%
      \expandafter\def\csname LTa\endcsname{\color{black}}%
      \expandafter\def\csname LT0\endcsname{\color[rgb]{1,0,0}}%
      \expandafter\def\csname LT1\endcsname{\color[rgb]{0,1,0}}%
      \expandafter\def\csname LT2\endcsname{\color[rgb]{0,0,1}}%
      \expandafter\def\csname LT3\endcsname{\color[rgb]{1,0,1}}%
      \expandafter\def\csname LT4\endcsname{\color[rgb]{0,1,1}}%
      \expandafter\def\csname LT5\endcsname{\color[rgb]{1,1,0}}%
      \expandafter\def\csname LT6\endcsname{\color[rgb]{0,0,0}}%
      \expandafter\def\csname LT7\endcsname{\color[rgb]{1,0.3,0}}%
      \expandafter\def\csname LT8\endcsname{\color[rgb]{0.5,0.5,0.5}}%
    \else
      \def\colorrgb#1{\color{black}}%
      \def\colorgray#1{\color[gray]{#1}}%
      \expandafter\def\csname LTw\endcsname{\color{white}}%
      \expandafter\def\csname LTb\endcsname{\color{black}}%
      \expandafter\def\csname LTa\endcsname{\color{black}}%
      \expandafter\def\csname LT0\endcsname{\color{black}}%
      \expandafter\def\csname LT1\endcsname{\color{black}}%
      \expandafter\def\csname LT2\endcsname{\color{black}}%
      \expandafter\def\csname LT3\endcsname{\color{black}}%
      \expandafter\def\csname LT4\endcsname{\color{black}}%
      \expandafter\def\csname LT5\endcsname{\color{black}}%
      \expandafter\def\csname LT6\endcsname{\color{black}}%
      \expandafter\def\csname LT7\endcsname{\color{black}}%
      \expandafter\def\csname LT8\endcsname{\color{black}}%
    \fi
  \fi
  \setlength{\unitlength}{0.0500bp}%
  \begin{picture}(4534.00,6802.00)%
    \gplgaddtomacro\gplbacktext{%
      \csname LTb\endcsname%
      \put(548,3537){\makebox(0,0)[r]{\strut{} 0}}%
      \put(548,3937){\makebox(0,0)[r]{\strut{} 10}}%
      \put(548,4337){\makebox(0,0)[r]{\strut{} 20}}%
      \put(548,4737){\makebox(0,0)[r]{\strut{} 30}}%
      \put(548,5137){\makebox(0,0)[r]{\strut{} 40}}%
      \put(548,5537){\makebox(0,0)[r]{\strut{} 50}}%
      \put(548,5937){\makebox(0,0)[r]{\strut{} 60}}%
      \put(548,6337){\makebox(0,0)[r]{\strut{} 70}}%
      \put(680,3317){\makebox(0,0){\strut{}}}%
      \put(1133,3317){\makebox(0,0){\strut{}}}%
      \put(1587,3317){\makebox(0,0){\strut{}}}%
      \put(2040,3317){\makebox(0,0){\strut{}}}%
      \put(2493,3317){\makebox(0,0){\strut{}}}%
      \put(2946,3317){\makebox(0,0){\strut{}}}%
      \put(3400,3317){\makebox(0,0){\strut{}}}%
      \put(3853,3317){\makebox(0,0){\strut{}}}%
      \put(4306,3317){\makebox(0,0){\strut{}}}%
      \put(1269,4137){\makebox(0,0)[l]{\strut{}$13.6$ [$nl$]}}%
      \put(2040,6337){\makebox(0,0)[l]{\strut{}$R_{C}$}}%
      \put(1451,6137){\rotatebox{90}{\makebox(0,0){\strut{}$R_{C}-W$}}}%
      \put(3490,6137){\rotatebox{90}{\makebox(0,0){\strut{}$R_{C}+W$}}}%
    }%
    \gplgaddtomacro\gplfronttext{%
      \csname LTb\endcsname%
      \put(566,6691){\makebox(0,0)[r]{\strut{}Dry}}%
      \csname LTb\endcsname%
      \put(1465,6691){\makebox(0,0)[r]{\strut{}Weig}}%
      \csname LTb\endcsname%
      \put(2364,6691){\makebox(0,0)[r]{\strut{}WilR}}%
      \csname LTb\endcsname%
      \put(3263,6691){\makebox(0,0)[r]{\strut{}WilF}}%
      \csname LTb\endcsname%
      \put(4162,6691){\makebox(0,0)[r]{\strut{}RabL}}%
    }%
    \gplgaddtomacro\gplbacktext{%
      \csname LTb\endcsname%
      \put(548,340){\makebox(0,0)[r]{\strut{} 0}}%
      \put(548,740){\makebox(0,0)[r]{\strut{} 10}}%
      \put(548,1140){\makebox(0,0)[r]{\strut{} 20}}%
      \put(548,1540){\makebox(0,0)[r]{\strut{} 30}}%
      \put(548,1941){\makebox(0,0)[r]{\strut{} 40}}%
      \put(548,2341){\makebox(0,0)[r]{\strut{} 50}}%
      \put(548,2741){\makebox(0,0)[r]{\strut{} 60}}%
      \put(548,3141){\makebox(0,0)[r]{\strut{} 70}}%
      \put(680,120){\makebox(0,0){\strut{}150}}%
      \put(1133,120){\makebox(0,0){\strut{}160}}%
      \put(1587,120){\makebox(0,0){\strut{}170}}%
      \put(2040,120){\makebox(0,0){\strut{}180}}%
      \put(2493,120){\makebox(0,0){\strut{}190}}%
      \put(2946,120){\makebox(0,0){\strut{}200}}%
      \put(3400,120){\makebox(0,0){\strut{}210}}%
      \put(3853,120){\makebox(0,0){\strut{}220}}%
      \put(4306,120){\makebox(0,0){\strut{}230}}%
      \put(174,3380){\rotatebox{-270}{\makebox(0,0){\strut{}$z$ [$mm$]}}}%
      \put(2493,-100){\makebox(0,0){\strut{}$r$ [$mm$]}}%
      \put(1269,940){\makebox(0,0)[l]{\strut{}$74.2$ [$nl$]}}%
    }%
    \gplgaddtomacro\gplfronttext{%
    }%
    \gplbacktext
    \put(0,0){\includegraphics{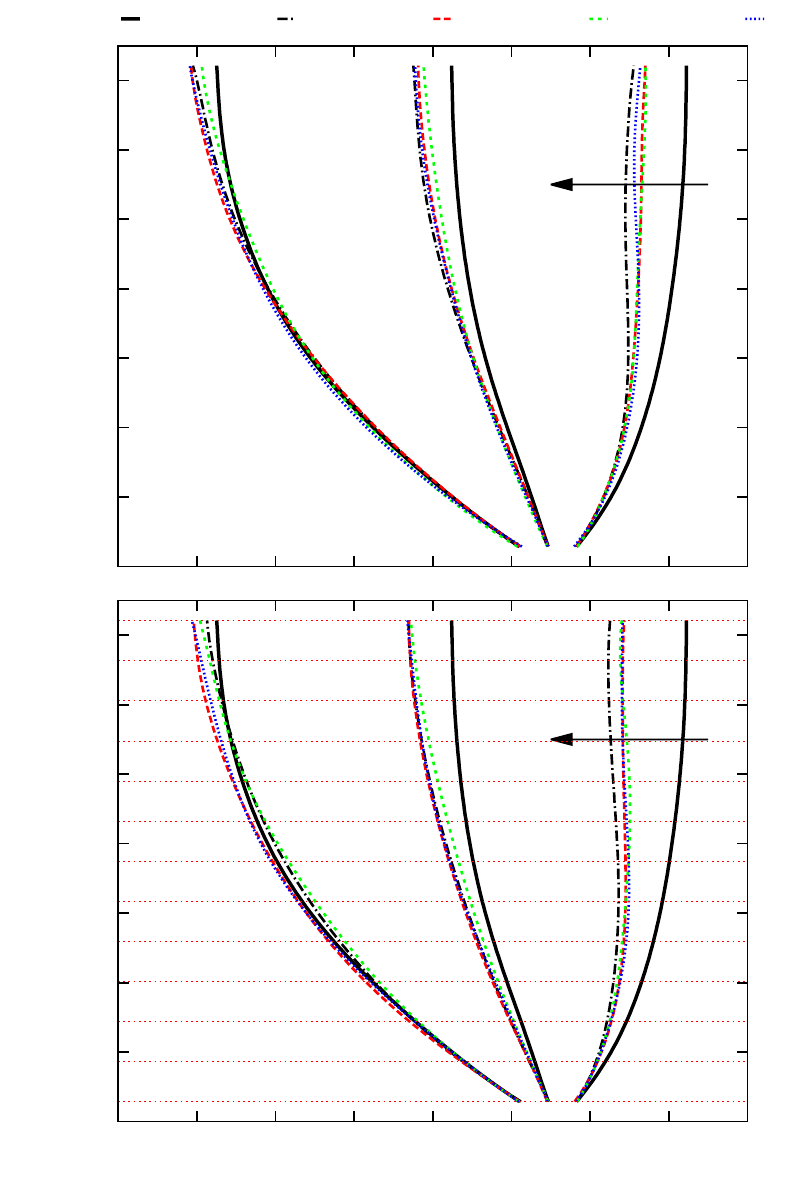}}%
    \gplfronttext
  \end{picture}%
\endgroup

%% file: TauP.tex
\begingroup
  \makeatletter
  \providecommand\color[2][]{%
    \GenericError{(gnuplot) \space\space\space\@spaces}{%
      Package color not loaded in conjunction with
      terminal option `colourtext'%
    }{See the gnuplot documentation for explanation.%
    }{Either use 'blacktext' in gnuplot or load the package
      color.sty in LaTeX.}%
    \renewcommand\color[2][]{}%
  }%
  \providecommand\includegraphics[2][]{%
    \GenericError{(gnuplot) \space\space\space\@spaces}{%
      Package graphicx or graphics not loaded%
    }{See the gnuplot documentation for explanation.%
    }{The gnuplot epslatex terminal needs graphicx.sty or graphics.sty.}%
    \renewcommand\includegraphics[2][]{}%
  }%
  \providecommand\rotatebox[2]{#2}%
  \@ifundefined{ifGPcolor}{%
    \newif\ifGPcolor
    \GPcolortrue
  }{}%
  \@ifundefined{ifGPblacktext}{%
    \newif\ifGPblacktext
    \GPblacktexttrue
  }{}%
  \let\gplgaddtomacro\g@addto@macro
  \gdef\gplbacktext{}%
  \gdef\gplfronttext{}%
  \makeatother
  \ifGPblacktext
    \def\colorrgb#1{}%
    \def\colorgray#1{}%
  \else
    \ifGPcolor
      \def\colorrgb#1{\color[rgb]{#1}}%
      \def\colorgray#1{\color[gray]{#1}}%
      \expandafter\def\csname LTw\endcsname{\color{white}}%
      \expandafter\def\csname LTb\endcsname{\color{black}}%
      \expandafter\def\csname LTa\endcsname{\color{black}}%
      \expandafter\def\csname LT0\endcsname{\color[rgb]{1,0,0}}%
      \expandafter\def\csname LT1\endcsname{\color[rgb]{0,1,0}}%
      \expandafter\def\csname LT2\endcsname{\color[rgb]{0,0,1}}%
      \expandafter\def\csname LT3\endcsname{\color[rgb]{1,0,1}}%
      \expandafter\def\csname LT4\endcsname{\color[rgb]{0,1,1}}%
      \expandafter\def\csname LT5\endcsname{\color[rgb]{1,1,0}}%
      \expandafter\def\csname LT6\endcsname{\color[rgb]{0,0,0}}%
      \expandafter\def\csname LT7\endcsname{\color[rgb]{1,0.3,0}}%
      \expandafter\def\csname LT8\endcsname{\color[rgb]{0.5,0.5,0.5}}%
    \else
      \def\colorrgb#1{\color{black}}%
      \def\colorgray#1{\color[gray]{#1}}%
      \expandafter\def\csname LTw\endcsname{\color{white}}%
      \expandafter\def\csname LTb\endcsname{\color{black}}%
      \expandafter\def\csname LTa\endcsname{\color{black}}%
      \expandafter\def\csname LT0\endcsname{\color{black}}%
      \expandafter\def\csname LT1\endcsname{\color{black}}%
      \expandafter\def\csname LT2\endcsname{\color{black}}%
      \expandafter\def\csname LT3\endcsname{\color{black}}%
      \expandafter\def\csname LT4\endcsname{\color{black}}%
      \expandafter\def\csname LT5\endcsname{\color{black}}%
      \expandafter\def\csname LT6\endcsname{\color{black}}%
      \expandafter\def\csname LT7\endcsname{\color{black}}%
      \expandafter\def\csname LT8\endcsname{\color{black}}%
    \fi
  \fi
  \setlength{\unitlength}{0.0500bp}%
  \begin{picture}(4534.00,6802.00)%
    \gplgaddtomacro\gplbacktext{%
      \csname LTb\endcsname%
      \put(548,3537){\makebox(0,0)[r]{\strut{} 0}}%
      \csname LTb\endcsname%
      \put(548,4137){\makebox(0,0)[r]{\strut{} 2}}%
      \csname LTb\endcsname%
      \put(548,4737){\makebox(0,0)[r]{\strut{} 4}}%
      \csname LTb\endcsname%
      \put(548,5337){\makebox(0,0)[r]{\strut{} 6}}%
      \csname LTb\endcsname%
      \put(548,5937){\makebox(0,0)[r]{\strut{} 8}}%
      \csname LTb\endcsname%
      \put(548,6537){\makebox(0,0)[r]{\strut{} 10}}%
      \csname LTb\endcsname%
      \put(680,3317){\makebox(0,0){\strut{}}}%
      \csname LTb\endcsname%
      \put(1133,3317){\makebox(0,0){\strut{}}}%
      \csname LTb\endcsname%
      \put(1587,3317){\makebox(0,0){\strut{}}}%
      \csname LTb\endcsname%
      \put(2040,3317){\makebox(0,0){\strut{}}}%
      \csname LTb\endcsname%
      \put(2493,3317){\makebox(0,0){\strut{}}}%
      \csname LTb\endcsname%
      \put(2946,3317){\makebox(0,0){\strut{}}}%
      \csname LTb\endcsname%
      \put(3400,3317){\makebox(0,0){\strut{}}}%
      \csname LTb\endcsname%
      \put(3853,3317){\makebox(0,0){\strut{}}}%
      \csname LTb\endcsname%
      \put(4306,3317){\makebox(0,0){\strut{}}}%
      \put(2946,3837){\makebox(0,0)[l]{\strut{}$13.6$ [$nl$]}}%
      \put(2221,4497){\rotatebox{41}{\makebox(0,0){\strut{}$0.13x$}}}%
      \put(1587,5187){\rotatebox{41}{\makebox(0,0){\strut{}$0.13x+1.3$}}}%
    }%
    \gplgaddtomacro\gplfronttext{%
      \csname LTb\endcsname%
      \put(566,6724){\makebox(0,0)[r]{\strut{}Dry}}%
      \csname LTb\endcsname%
      \put(1465,6724){\makebox(0,0)[r]{\strut{}Weig}}%
      \csname LTb\endcsname%
      \put(2364,6724){\makebox(0,0)[r]{\strut{}WilR}}%
      \csname LTb\endcsname%
      \put(3263,6724){\makebox(0,0)[r]{\strut{}WilF}}%
      \csname LTb\endcsname%
      \put(4162,6724){\makebox(0,0)[r]{\strut{}RabL}}%
    }%
    \gplgaddtomacro\gplbacktext{%
      \csname LTb\endcsname%
      \put(548,340){\makebox(0,0)[r]{\strut{} 0}}%
      \csname LTb\endcsname%
      \put(548,940){\makebox(0,0)[r]{\strut{} 2}}%
      \csname LTb\endcsname%
      \put(548,1540){\makebox(0,0)[r]{\strut{} 4}}%
      \csname LTb\endcsname%
      \put(548,2141){\makebox(0,0)[r]{\strut{} 6}}%
      \csname LTb\endcsname%
      \put(548,2741){\makebox(0,0)[r]{\strut{} 8}}%
      \csname LTb\endcsname%
      \put(548,3341){\makebox(0,0)[r]{\strut{} 10}}%
      \csname LTb\endcsname%
      \put(680,120){\makebox(0,0){\strut{}0}}%
      \csname LTb\endcsname%
      \put(1133,120){\makebox(0,0){\strut{}10}}%
      \csname LTb\endcsname%
      \put(1587,120){\makebox(0,0){\strut{}20}}%
      \csname LTb\endcsname%
      \put(2040,120){\makebox(0,0){\strut{}30}}%
      \csname LTb\endcsname%
      \put(2493,120){\makebox(0,0){\strut{}40}}%
      \csname LTb\endcsname%
      \put(2946,120){\makebox(0,0){\strut{}50}}%
      \csname LTb\endcsname%
      \put(3400,120){\makebox(0,0){\strut{}60}}%
      \csname LTb\endcsname%
      \put(3853,120){\makebox(0,0){\strut{}70}}%
      \csname LTb\endcsname%
      \put(4306,120){\makebox(0,0){\strut{}80}}%
      \put(174,3380){\rotatebox{-270}{\makebox(0,0){\strut{}$\tau$ [$Pa$]}}}%
      \put(2493,-100){\makebox(0,0){\strut{}$p$ [$Pa$]}}%
      \put(2946,640){\makebox(0,0)[l]{\strut{}$74.2$ [$nl$]}}%
      \put(2221,1300){\rotatebox{41}{\makebox(0,0){\strut{}$0.13x$}}}%
      \put(1587,2231){\rotatebox{41}{\makebox(0,0){\strut{}$0.13x+2.0$}}}%
    }%
    \gplgaddtomacro\gplfronttext{%
    }%
    \gplbacktext
    \put(0,0){\includegraphics{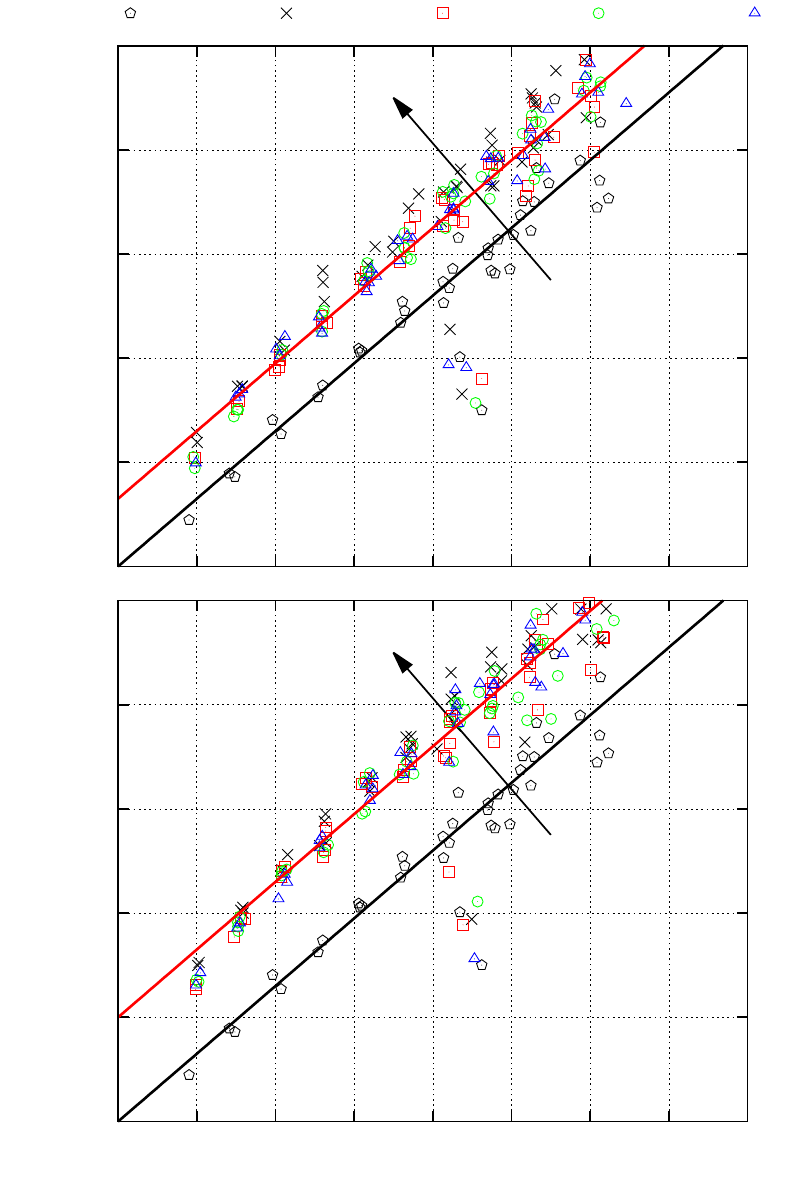}}%
    \gplfronttext
  \end{picture}%
\endgroup

%% file: torque_vergl.tex
\begingroup
  \makeatletter
  \providecommand\color[2][]{%
    \GenericError{(gnuplot) \space\space\space\@spaces}{%
      Package color not loaded in conjunction with
      terminal option `colourtext'%
    }{See the gnuplot documentation for explanation.%
    }{Either use 'blacktext' in gnuplot or load the package
      color.sty in LaTeX.}%
    \renewcommand\color[2][]{}%
  }%
  \providecommand\includegraphics[2][]{%
    \GenericError{(gnuplot) \space\space\space\@spaces}{%
      Package graphicx or graphics not loaded%
    }{See the gnuplot documentation for explanation.%
    }{The gnuplot epslatex terminal needs graphicx.sty or graphics.sty.}%
    \renewcommand\includegraphics[2][]{}%
  }%
  \providecommand\rotatebox[2]{#2}%
  \@ifundefined{ifGPcolor}{%
    \newif\ifGPcolor
    \GPcolortrue
  }{}%
  \@ifundefined{ifGPblacktext}{%
    \newif\ifGPblacktext
    \GPblacktexttrue
  }{}%
  \let\gplgaddtomacro\g@addto@macro
  \gdef\gplbacktext{}%
  \gdef\gplfronttext{}%
  \makeatother
  \ifGPblacktext
    \def\colorrgb#1{}%
    \def\colorgray#1{}%
  \else
    \ifGPcolor
      \def\colorrgb#1{\color[rgb]{#1}}%
      \def\colorgray#1{\color[gray]{#1}}%
      \expandafter\def\csname LTw\endcsname{\color{white}}%
      \expandafter\def\csname LTb\endcsname{\color{black}}%
      \expandafter\def\csname LTa\endcsname{\color{black}}%
      \expandafter\def\csname LT0\endcsname{\color[rgb]{1,0,0}}%
      \expandafter\def\csname LT1\endcsname{\color[rgb]{0,1,0}}%
      \expandafter\def\csname LT2\endcsname{\color[rgb]{0,0,1}}%
      \expandafter\def\csname LT3\endcsname{\color[rgb]{1,0,1}}%
      \expandafter\def\csname LT4\endcsname{\color[rgb]{0,1,1}}%
      \expandafter\def\csname LT5\endcsname{\color[rgb]{1,1,0}}%
      \expandafter\def\csname LT6\endcsname{\color[rgb]{0,0,0}}%
      \expandafter\def\csname LT7\endcsname{\color[rgb]{1,0.3,0}}%
      \expandafter\def\csname LT8\endcsname{\color[rgb]{0.5,0.5,0.5}}%
    \else
      \def\colorrgb#1{\color{black}}%
      \def\colorgray#1{\color[gray]{#1}}%
      \expandafter\def\csname LTw\endcsname{\color{white}}%
      \expandafter\def\csname LTb\endcsname{\color{black}}%
      \expandafter\def\csname LTa\endcsname{\color{black}}%
      \expandafter\def\csname LT0\endcsname{\color{black}}%
      \expandafter\def\csname LT1\endcsname{\color{black}}%
      \expandafter\def\csname LT2\endcsname{\color{black}}%
      \expandafter\def\csname LT3\endcsname{\color{black}}%
      \expandafter\def\csname LT4\endcsname{\color{black}}%
      \expandafter\def\csname LT5\endcsname{\color{black}}%
      \expandafter\def\csname LT6\endcsname{\color{black}}%
      \expandafter\def\csname LT7\endcsname{\color{black}}%
      \expandafter\def\csname LT8\endcsname{\color{black}}%
    \fi
  \fi
  \setlength{\unitlength}{0.0500bp}%
  \begin{picture}(4762.00,2380.00)%
    \gplgaddtomacro\gplbacktext{%
      \csname LTb\endcsname%
      \put(858,264){\makebox(0,0)[r]{\strut{} 1.3}}%
      \csname LTb\endcsname%
      \put(858,573){\makebox(0,0)[r]{\strut{} 1.35}}%
      \csname LTb\endcsname%
      \put(858,881){\makebox(0,0)[r]{\strut{} 1.4}}%
      \csname LTb\endcsname%
      \put(858,1190){\makebox(0,0)[r]{\strut{} 1.45}}%
      \csname LTb\endcsname%
      \put(858,1498){\makebox(0,0)[r]{\strut{} 1.5}}%
      \csname LTb\endcsname%
      \put(858,1807){\makebox(0,0)[r]{\strut{} 1.55}}%
      \csname LTb\endcsname%
      \put(858,2115){\makebox(0,0)[r]{\strut{} 1.6}}%
      \put(352,1189){\rotatebox{-270}{\makebox(0,0){\strut{}$\vec{M}$ [$Nm$]}}}%
    }%
    \gplgaddtomacro\gplfronttext{%
      \csname LTb\endcsname%
      \put(2065,2221){\makebox(0,0)[r]{\strut{}$V=13.6$ [$nl$]}}%
      \csname LTb\endcsname%
      \put(3901,2221){\makebox(0,0)[r]{\strut{}$V=74.2$ [$nl$]}}%
      \csname LTb\endcsname%
      \put(1328,-43){\rotatebox{60}{\makebox(0,0){\strut{}Dry}}}%
      \put(2003,-43){\rotatebox{60}{\makebox(0,0){\strut{}Weig}}}%
      \put(2678,-43){\rotatebox{60}{\makebox(0,0){\strut{}WilR}}}%
      \put(3353,-43){\rotatebox{60}{\makebox(0,0){\strut{}WilF}}}%
      \put(4028,-43){\rotatebox{60}{\makebox(0,0){\strut{}RabL}}}%
    }%
    \gplbacktext
    \put(0,0){\includegraphics{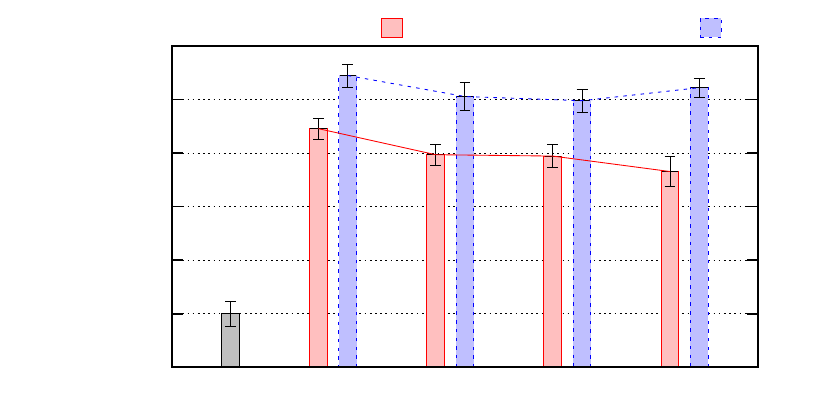}}%
    \gplfronttext
  \end{picture}%
\endgroup

%% file: contactsA.tex
\begingroup
  \makeatletter
  \providecommand\color[2][]{%
    \GenericError{(gnuplot) \space\space\space\@spaces}{%
      Package color not loaded in conjunction with
      terminal option `colourtext'%
    }{See the gnuplot documentation for explanation.%
    }{Either use 'blacktext' in gnuplot or load the package
      color.sty in LaTeX.}%
    \renewcommand\color[2][]{}%
  }%
  \providecommand\includegraphics[2][]{%
    \GenericError{(gnuplot) \space\space\space\@spaces}{%
      Package graphicx or graphics not loaded%
    }{See the gnuplot documentation for explanation.%
    }{The gnuplot epslatex terminal needs graphicx.sty or graphics.sty.}%
    \renewcommand\includegraphics[2][]{}%
  }%
  \providecommand\rotatebox[2]{#2}%
  \@ifundefined{ifGPcolor}{%
    \newif\ifGPcolor
    \GPcolortrue
  }{}%
  \@ifundefined{ifGPblacktext}{%
    \newif\ifGPblacktext
    \GPblacktexttrue
  }{}%
  \let\gplgaddtomacro\g@addto@macro
  \gdef\gplbacktext{}%
  \gdef\gplfronttext{}%
  \makeatother
  \ifGPblacktext
    \def\colorrgb#1{}%
    \def\colorgray#1{}%
  \else
    \ifGPcolor
      \def\colorrgb#1{\color[rgb]{#1}}%
      \def\colorgray#1{\color[gray]{#1}}%
      \expandafter\def\csname LTw\endcsname{\color{white}}%
      \expandafter\def\csname LTb\endcsname{\color{black}}%
      \expandafter\def\csname LTa\endcsname{\color{black}}%
      \expandafter\def\csname LT0\endcsname{\color[rgb]{1,0,0}}%
      \expandafter\def\csname LT1\endcsname{\color[rgb]{0,1,0}}%
      \expandafter\def\csname LT2\endcsname{\color[rgb]{0,0,1}}%
      \expandafter\def\csname LT3\endcsname{\color[rgb]{1,0,1}}%
      \expandafter\def\csname LT4\endcsname{\color[rgb]{0,1,1}}%
      \expandafter\def\csname LT5\endcsname{\color[rgb]{1,1,0}}%
      \expandafter\def\csname LT6\endcsname{\color[rgb]{0,0,0}}%
      \expandafter\def\csname LT7\endcsname{\color[rgb]{1,0.3,0}}%
      \expandafter\def\csname LT8\endcsname{\color[rgb]{0.5,0.5,0.5}}%
    \else
      \def\colorrgb#1{\color{black}}%
      \def\colorgray#1{\color[gray]{#1}}%
      \expandafter\def\csname LTw\endcsname{\color{white}}%
      \expandafter\def\csname LTb\endcsname{\color{black}}%
      \expandafter\def\csname LTa\endcsname{\color{black}}%
      \expandafter\def\csname LT0\endcsname{\color{black}}%
      \expandafter\def\csname LT1\endcsname{\color{black}}%
      \expandafter\def\csname LT2\endcsname{\color{black}}%
      \expandafter\def\csname LT3\endcsname{\color{black}}%
      \expandafter\def\csname LT4\endcsname{\color{black}}%
      \expandafter\def\csname LT5\endcsname{\color{black}}%
      \expandafter\def\csname LT6\endcsname{\color{black}}%
      \expandafter\def\csname LT7\endcsname{\color{black}}%
      \expandafter\def\csname LT8\endcsname{\color{black}}%
    \fi
  \fi
  \setlength{\unitlength}{0.0500bp}%
  \begin{picture}(4534.00,6802.00)%
    \gplgaddtomacro\gplbacktext{%
      \csname LTb\endcsname%
      \put(94,3872){\makebox(0,0)[r]{\strut{}$10^{1}$}}%
      \put(94,4450){\makebox(0,0)[r]{\strut{}$10^{2}$}}%
      \put(94,5029){\makebox(0,0)[r]{\strut{}$10^{3}$}}%
      \put(94,5607){\makebox(0,0)[r]{\strut{}$10^{4}$}}%
      \put(94,6185){\makebox(0,0)[r]{\strut{}$10^{5}$}}%
      \put(226,3248){\makebox(0,0){\strut{}}}%
      \put(770,3248){\makebox(0,0){\strut{}}}%
      \put(380,4964){\rotatebox{-270}{\makebox(0,0){\strut{}$N$ [$-$]}}}%
    }%
    \gplgaddtomacro\gplfronttext{%
    }%
    \gplgaddtomacro\gplbacktext{%
      \csname LTb\endcsname%
      \put(1405,3248){\makebox(0,0){\strut{}}}%
      \put(2050,3248){\makebox(0,0){\strut{}}}%
      \put(2694,3248){\makebox(0,0){\strut{}}}%
      \put(3339,3248){\makebox(0,0){\strut{}}}%
      \put(3984,3248){\makebox(0,0){\strut{}}}%
      \put(4438,3872){\makebox(0,0)[l]{\strut{}$10^{1}$}}%
      \put(4438,4450){\makebox(0,0)[l]{\strut{}$10^{2}$}}%
      \put(4438,5029){\makebox(0,0)[l]{\strut{}$10^{3}$}}%
      \put(4438,5607){\makebox(0,0)[l]{\strut{}$10^{4}$}}%
      \put(4438,6185){\makebox(0,0)[l]{\strut{}$10^{5}$}}%
      \put(4138,4964){\rotatebox{-270}{\makebox(0,0){\strut{}$N$ [$-$]}}}%
    }%
    \gplgaddtomacro\gplfronttext{%
      \csname LTb\endcsname%
      \put(3552,4081){\makebox(0,0)[r]{\strut{}Dry}}%
      \csname LTb\endcsname%
      \put(3552,3861){\makebox(0,0)[r]{\strut{}Weig}}%
      \csname LTb\endcsname%
      \put(3552,3641){\makebox(0,0)[r]{\strut{}WilF}}%
    }%
    \gplgaddtomacro\gplbacktext{%
      \csname LTb\endcsname%
      \put(94,767){\makebox(0,0)[r]{\strut{}-12000}}%
      \put(94,1195){\makebox(0,0)[r]{\strut{}-9000}}%
      \put(94,1622){\makebox(0,0)[r]{\strut{}-6000}}%
      \put(94,2050){\makebox(0,0)[r]{\strut{}-3000}}%
      \put(94,2477){\makebox(0,0)[r]{\strut{}0}}%
      \put(94,2905){\makebox(0,0)[r]{\strut{}3000}}%
      \put(226,120){\makebox(0,0){\strut{}-80}}%
      \put(770,120){\makebox(0,0){\strut{}-40}}%
      \put(1314,120){\makebox(0,0){\strut{}0}}%
      \put(380,1836){\rotatebox{-270}{\makebox(0,0){\strut{}$F_{norm}$ [$\mu N$]}}}%
    }%
    \gplgaddtomacro\gplfronttext{%
    }%
    \gplgaddtomacro\gplbacktext{%
      \csname LTb\endcsname%
      \put(2050,120){\makebox(0,0){\strut{}100}}%
      \put(2694,120){\makebox(0,0){\strut{}200}}%
      \put(3339,120){\makebox(0,0){\strut{}300}}%
      \put(3984,120){\makebox(0,0){\strut{}400}}%
      \put(4438,340){\makebox(0,0)[l]{\strut{}0}}%
      \put(4438,767){\makebox(0,0)[l]{\strut{}50}}%
      \put(4438,1195){\makebox(0,0)[l]{\strut{}100}}%
      \put(4438,1622){\makebox(0,0)[l]{\strut{}150}}%
      \put(4438,2050){\makebox(0,0)[l]{\strut{}200}}%
      \put(4438,2477){\makebox(0,0)[l]{\strut{}250}}%
      \put(4438,2905){\makebox(0,0)[l]{\strut{}300}}%
      \put(4438,3332){\makebox(0,0)[l]{\strut{}350}}%
      \put(4151,1836){\rotatebox{-270}{\makebox(0,0){\strut{}${F}_{cap}$ [$\mu N$]}}}%
      \put(3647,450){\makebox(0,0){\strut{}$a$ [$\mu m$]}}%
    }%
    \gplgaddtomacro\gplfronttext{%
    }%
    \gplgaddtomacro\gplbacktext{%
      \csname LTb\endcsname%
      \put(2629,5153){\makebox(0,0){\strut{}-10}}%
      \put(3128,5153){\makebox(0,0){\strut{}0}}%
      \put(3626,5153){\makebox(0,0){\strut{}10}}%
      \put(3758,5373){\makebox(0,0)[l]{\strut{}$10^{3}$}}%
      \put(3758,5757){\makebox(0,0)[l]{\strut{}$10^{4}$}}%
      \put(3758,6141){\makebox(0,0)[l]{\strut{}$10^{5}$}}%
    }%
    \gplgaddtomacro\gplfronttext{%
    }%
    \gplgaddtomacro\gplbacktext{%
      \csname LTb\endcsname%
      \put(2413,1820){\makebox(0,0){\strut{}0}}%
      \put(2977,1820){\makebox(0,0){\strut{}20}}%
      \put(3531,2040){\makebox(0,0)[l]{\strut{}220}}%
      \put(3531,2312){\makebox(0,0)[l]{\strut{}240}}%
      \put(3531,2584){\makebox(0,0)[l]{\strut{}260}}%
      \put(3531,2856){\makebox(0,0)[l]{\strut{}280}}%
    }%
    \gplgaddtomacro\gplfronttext{%
    }%
    \gplbacktext
    \put(0,0){\includegraphics{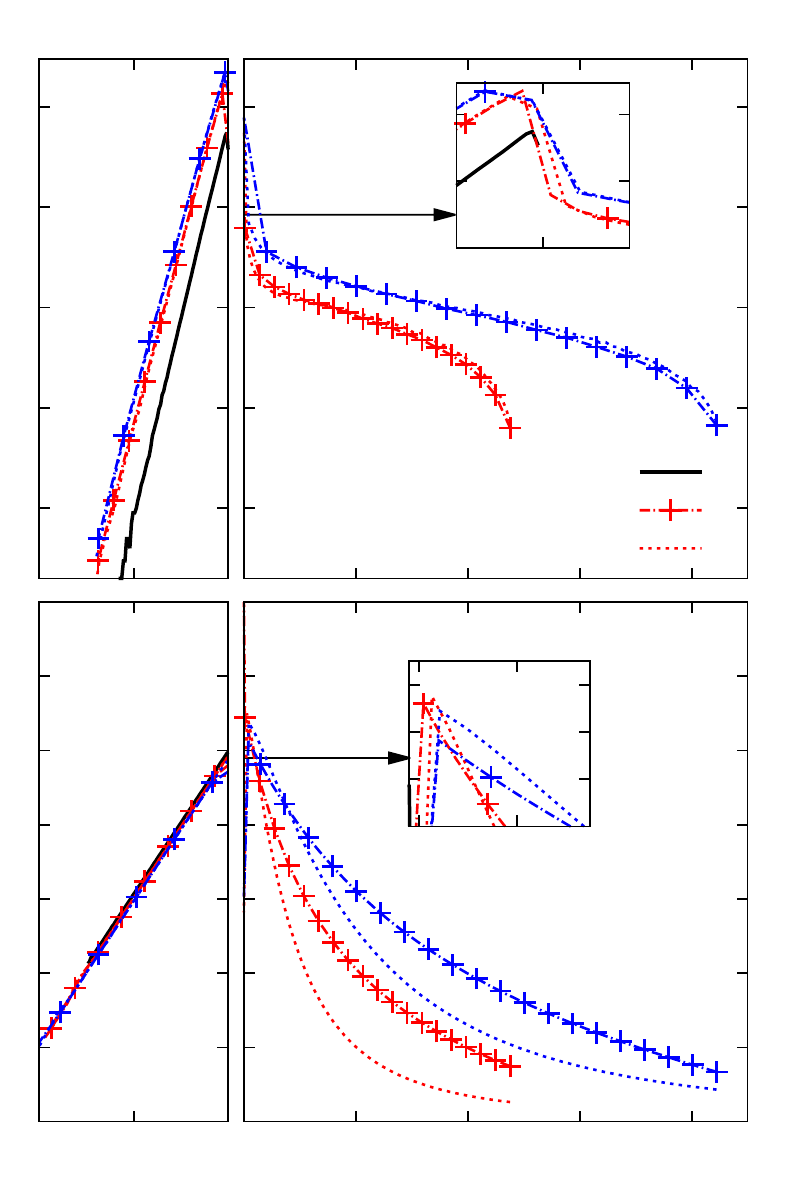}}%
    \gplfronttext
  \end{picture}%
\endgroup

%% file: ContactVert.tex
\begingroup
  \makeatletter
  \providecommand\color[2][]{%
    \GenericError{(gnuplot) \space\space\space\@spaces}{%
      Package color not loaded in conjunction with
      terminal option `colourtext'%
    }{See the gnuplot documentation for explanation.%
    }{Either use 'blacktext' in gnuplot or load the package
      color.sty in LaTeX.}%
    \renewcommand\color[2][]{}%
  }%
  \providecommand\includegraphics[2][]{%
    \GenericError{(gnuplot) \space\space\space\@spaces}{%
      Package graphicx or graphics not loaded%
    }{See the gnuplot documentation for explanation.%
    }{The gnuplot epslatex terminal needs graphicx.sty or graphics.sty.}%
    \renewcommand\includegraphics[2][]{}%
  }%
  \providecommand\rotatebox[2]{#2}%
  \@ifundefined{ifGPcolor}{%
    \newif\ifGPcolor
    \GPcolortrue
  }{}%
  \@ifundefined{ifGPblacktext}{%
    \newif\ifGPblacktext
    \GPblacktexttrue
  }{}%
  \let\gplgaddtomacro\g@addto@macro
  \gdef\gplbacktext{}%
  \gdef\gplfronttext{}%
  \makeatother
  \ifGPblacktext
    \def\colorrgb#1{}%
    \def\colorgray#1{}%
  \else
    \ifGPcolor
      \def\colorrgb#1{\color[rgb]{#1}}%
      \def\colorgray#1{\color[gray]{#1}}%
      \expandafter\def\csname LTw\endcsname{\color{white}}%
      \expandafter\def\csname LTb\endcsname{\color{black}}%
      \expandafter\def\csname LTa\endcsname{\color{black}}%
      \expandafter\def\csname LT0\endcsname{\color[rgb]{1,0,0}}%
      \expandafter\def\csname LT1\endcsname{\color[rgb]{0,1,0}}%
      \expandafter\def\csname LT2\endcsname{\color[rgb]{0,0,1}}%
      \expandafter\def\csname LT3\endcsname{\color[rgb]{1,0,1}}%
      \expandafter\def\csname LT4\endcsname{\color[rgb]{0,1,1}}%
      \expandafter\def\csname LT5\endcsname{\color[rgb]{1,1,0}}%
      \expandafter\def\csname LT6\endcsname{\color[rgb]{0,0,0}}%
      \expandafter\def\csname LT7\endcsname{\color[rgb]{1,0.3,0}}%
      \expandafter\def\csname LT8\endcsname{\color[rgb]{0.5,0.5,0.5}}%
    \else
      \def\colorrgb#1{\color{black}}%
      \def\colorgray#1{\color[gray]{#1}}%
      \expandafter\def\csname LTw\endcsname{\color{white}}%
      \expandafter\def\csname LTb\endcsname{\color{black}}%
      \expandafter\def\csname LTa\endcsname{\color{black}}%
      \expandafter\def\csname LT0\endcsname{\color{black}}%
      \expandafter\def\csname LT1\endcsname{\color{black}}%
      \expandafter\def\csname LT2\endcsname{\color{black}}%
      \expandafter\def\csname LT3\endcsname{\color{black}}%
      \expandafter\def\csname LT4\endcsname{\color{black}}%
      \expandafter\def\csname LT5\endcsname{\color{black}}%
      \expandafter\def\csname LT6\endcsname{\color{black}}%
      \expandafter\def\csname LT7\endcsname{\color{black}}%
      \expandafter\def\csname LT8\endcsname{\color{black}}%
    \fi
  \fi
  \setlength{\unitlength}{0.0500bp}%
  \begin{picture}(4762.00,3514.00)%
    \gplgaddtomacro\gplbacktext{%
      \csname LTb\endcsname%
      \put(344,351){\makebox(0,0)[r]{\strut{}$10^{2}$}}%
      \csname LTb\endcsname%
      \put(344,834){\makebox(0,0)[r]{\strut{}$10^{3}$}}%
      \csname LTb\endcsname%
      \put(344,1317){\makebox(0,0)[r]{\strut{}$10^{4}$}}%
      \csname LTb\endcsname%
      \put(344,1800){\makebox(0,0)[r]{\strut{}$10^{5}$}}%
      \csname LTb\endcsname%
      \put(344,2283){\makebox(0,0)[r]{\strut{}$10^{6}$}}%
      \csname LTb\endcsname%
      \put(344,2766){\makebox(0,0)[r]{\strut{}$10^{7}$}}%
      \csname LTb\endcsname%
      \put(344,3249){\makebox(0,0)[r]{\strut{}$10^{8}$}}%
      \csname LTb\endcsname%
      \put(476,131){\makebox(0,0){\strut{}-100}}%
      \csname LTb\endcsname%
      \put(1150,131){\makebox(0,0){\strut{} 0}}%
      \csname LTb\endcsname%
      \put(1825,131){\makebox(0,0){\strut{} 100}}%
      \csname LTb\endcsname%
      \put(2499,131){\makebox(0,0){\strut{} 200}}%
      \csname LTb\endcsname%
      \put(3173,131){\makebox(0,0){\strut{} 300}}%
      \csname LTb\endcsname%
      \put(3848,131){\makebox(0,0){\strut{} 400}}%
      \csname LTb\endcsname%
      \put(4522,131){\makebox(0,0){\strut{} 500}}%
      \put(696,1800){\rotatebox{-270}{\makebox(0,0){\strut{}$n  [m^{-3}]$}}}%
      \put(2499,-89){\makebox(0,0){\strut{}$a$ [$\mu m$]}}%
    }%
    \gplgaddtomacro\gplfronttext{%
      \csname LTb\endcsname%
      \put(3504,3076){\makebox(0,0)[r]{\strut{}Inside band}}%
      \csname LTb\endcsname%
      \put(3504,2856){\makebox(0,0)[r]{\strut{}Outside band}}%
    }%
    \gplbacktext
    \put(0,0){\includegraphics{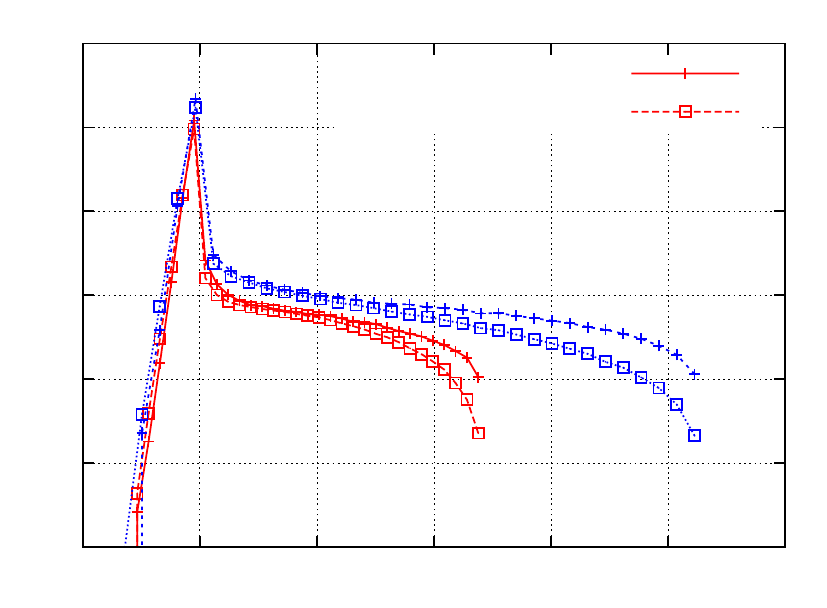}}%
    \gplfronttext
  \end{picture}%
\endgroup

%% file: appendix.tex
\begin{appendix}
\renewcommand{\theequation}{\thesection\arabic{equation}}
\setcounter{equation}{0}
\section{Weigert's model equations}
\label{WeigertAp}
The following equations are used for the calculation of $\beta$, $C_a$ and $C_{\theta}$,
$p_k$, $R_1$ and $R_2$ for Weigert's model.
$V$ (bridge volume) can be found as:
      \begin{equation}
        \label{eq:WEI:15}
        V = 0.12 {(2R)}^3 \sin^4{\beta} C_a C_{\theta}
      \end{equation}

Since $V$ is a given parameter for us, $\beta$ can be obtained:
      \begin{equation}
        \label{eq:WEI:15Mod}
        \beta = \arcsin{\left( \cfrac{V}{0.12 \cdot (2R)^3 \cdot C_a C_{\theta}} \right)^{1/4}}
      \end{equation}
$C_a$ and $C_{\theta}$ are the correction functions for the distance and the contact   
angles respectively, they are calculated according to:
      \begin{equation}
        \label{eq:WEI:16}
        C_{a} = \left( 1 + 6 \cfrac{a}{2R} \right)
      \end{equation}
      \begin{equation}
        \label{eq:WEI:17}
        C_{\theta} = \left( 1 + 1.1 \cdot \sin{\theta} \right)
      \end{equation}
The capillary pressure $p_k$ is calculated according to the Laplace-Young equation:
      \begin{equation}
        \label{eq:WEI:22}
        p_k = \gamma\left( \cfrac{1}{R_1} + \cfrac{1}{R_2} \right)
      \end{equation}
    
The principal radii of the bridge curvature $R_1$ and $R_2$ are taken 
positive and negative respectively, and calculated according to Pietsch and Rumpf 
\cite{Pietsch1967}:
      \begin{equation}
        \label{eq:Piet:2}
        R_1 = \cfrac{R (1-\cos{\beta}) + a }{\cos{(\beta + \theta)}}
      \end{equation}
      \begin{equation}
        \label{eq:Piet:3}
        R_2 = R \sin{\beta} + R_1\left[ \sin(\beta + \theta) -1\right]
      \end{equation}
The full description of the model can be found here \cite{Weigert1999}.

\section{Willett's full model equations}
\setcounter{equation}{0}
The coefficients $f_{1\dots 4}$ of Willett's full model are calculated as follows: 
      \begin{equation}
        \begin{split}
        \label{eq:WIL:A1-f1}
          f_1 = &(-0.44507 + 0.050832\ \theta - 1.1466\ {\theta}^2) + \\
                &(-0.1119 - 0.000411\ \theta - 0.1490\ {\theta}^2) \  \ln(V^{*}) + \\
                &(-0.012101 - 0.0036456\ \theta - 0.01255\ {\theta}^2) \ {(\ln(V^{*}))}^2 +\\
                &(-0.0005 - 0.0003505\ \theta - 0.00029076\ {\theta}^2) \ {(\ln(V^{*}))}^3
        \end{split}
      \end{equation}
      \begin{equation}
        \begin{split}
        \label{eq:WIL:A1-f2}
        f2 = &(1.9222 - 0.57473\ \theta - 1.2918\ {\theta}^2) + \\
             &(-0.0668 - 0.1201\ \theta - 0.22574\ {\theta}^2) \  \ln(V^{*}) +\\
             &(-0.0013375 - 0.0068988\ \theta - 0.01137\ {\theta}^2) \ {(\ln(V^{*}))}^2
        \end{split}
      \end{equation}
      \begin{equation}
        \begin{split}
        \label{eq:WIL:A1-f3}
        f3 = &(1.268 - 0.01396\ \theta - 0.23566\ {\theta}^2) + \\
             &(0.198 + 0.092\ \theta - 0.06418\ {\theta}^2) \  \ln(V^{*}) +\\
             &(0.02232 + 0.02238\ \theta - 0.009853\ {\theta}^2) \ {(\ln(V^{*}))}^2 +\\
             &(0.0008585 + 0.001318\ \theta - 0.00053\ {\theta}^2) \ {(\ln(V^{*}))}^3;
        \end{split}
      \end{equation}
      \begin{equation}
        \begin{split}
        \label{eq:WIL:A1-f4}
        f4 = &(-0.010703 + 0.073776\ \theta - 0.34742\ {\theta}^2) +\\
             &(0.03345 + 0.04543\ \theta - 0.09056\ {\theta}^2) \  \ln(V^{*}) +\\
             &(0.0018574 + 0.004456\ \theta - 0.006257\ {\theta}^2) \ {(\ln(V^{*}))}^2;
        \end{split}
      \end{equation}
The full description of the model can be found here \cite{Willett2000}.

\end{appendix}